\documentclass[journal]{IEEEtran}
\IEEEoverridecommandlockouts
\usepackage{cite}
 \usepackage[utf8]{inputenc}
     \usepackage{setspace}
\usepackage{amsmath,amssymb,amsfonts}
\usepackage{dirtytalk}
\usepackage{mathrsfs}
\usepackage{algorithmic}
\usepackage{graphicx}
\usepackage{textcomp}
\usepackage{xcolor}
\usepackage{longtable}
\usepackage{colortbl}
\usepackage{amsfonts}
\DeclareMathOperator{\sinc}{sinc}
\usepackage{etoolbox}
\makeatletter
\usepackage{enumitem}
\usepackage{tabu}
\usepackage{array} 

\usepackage{soul}

\patchcmd{\@makecaption}
  {\scshape}
  {}
  {}
  {}
\makeatletter
\patchcmd{\@makecaption}
  {\\}
  {.\ }
  {}
  {}
\makeatother
\def\tablename{Table}

\usepackage{siunitx}
\usepackage{gensymb}
\usepackage{bbm}
\usepackage{algorithm} 
\usepackage{amsmath}
\DeclareMathOperator*{\argmax}{arg\,max}
\DeclareMathOperator*{\argmin}{arg\,min}
\usepackage{multirow}
\usepackage[english]{babel} 
\usepackage{blindtext}
\newcommand{\RNum}[1]{\uppercase\expandafter{\romannumeral #1\relax}}
\usepackage{scalerel,stackengine,amsmath}
\usepackage{atbegshi}
\AtBeginDocument{\AtBeginShipoutNext{\AtBeginShipoutDiscard}}
\usepackage{scalerel}
\usepackage{stackengine,wasysym}

\def\mathcolor#1#{\@mathcolor{#1}}
\def\@mathcolor#1#2#3{%
  \protect\leavevmode
  \begingroup
    \color#1{#2}#3%
  \endgroup
}
\usepackage[cmintegrals]{newtxmath}

\begin{document}

\title{Learning-to-Learn the Wave Angle Estimation}






{\author{Eray~Güven,~\IEEEmembership{Student Member,~IEEE,} Güneş~Karabulut Kurt,~\IEEEmembership{Senior Member,~IEEE}
}
\thanks{This work was supported in part by the Natural Sciences and Engineering Research Council of Canada (NSERC) Discovery Grant program.}
\thanks{E. Guven and G. Karabulut Kurt are  with the {Department of Electrical Engineering, Polytechnique Montr\'eal, Montr\'eal, Canada, e-mail: \{guven.eray, gunes.kurt\}@polymtl.ca}}
}

\maketitle

\begin{abstract}
A precise incident wave angle estimation in aerial communication is a key enabler in sixth-generation wireless communication network. With this goal, a generic 3-dimensional (3D) channel model is analyzed for air-to-air (A2A) networks under antenna misalignment, radio frequency impairments and polarization loss. The unique aspects of each aerial node are highlighted and the few-shot learning as a model agnostic meta-learning (MAML) classifier is proposed for learning-to-learn (L2L) incident wave angle estimation by utilizing the received signal strength (RSS). Additionally, a more computationally efficient technique, first order model agnostic meta-learning (FOMAML) is implemented. It has been observed that the proposed approach reaches up to 85\% training accuracy and 75.4\% evaluation accuracy with MAML. Regarding this, a convergence rate and accuracy trade-off have been established for several cases of MAML and FOMAML. For different L2L models trained with limited data, heuristic accuracy performance is determined by an upper bound of the probability of confidence. 
\end{abstract}

\setcounter{page}{1}
\section{Introduction}
Air-to-air (A2A) networks will play a crucial role in sixth-
generation (6G) wireless communication networks. \textcolor{black}{A2A enables continuous connectivity for aerial nodes over a wide area, leveraging line-of-sight (LoS) links, prompting the necessity of three-dimensional (3D) channel models for A2A networks for accurate performance evaluations \cite{goddemeier2015investigation}.} Among the stochastic and deterministic 3D channel models, the impact of the antenna gain over the channel is an open issue that changes according to azimuth and elevation. We emphasize that assumptions of \textcolor{black}{the} ideal antenna and purely stable dynamic aerial node make the 3D channel models for A2A networks inapplicable. With a fixed on-board dipole antenna, a dynamic aerial node leads to dynamic antenna orientation. Under the antenna defects, polarization loss and antenna misalignment, \textcolor{black}{the} 3D channel model becomes more complex than 2-dimensional idealized channel models due to the spatial characteristic from the mobility and the additional dimension. \textcolor{black}{Therefore, inherent 3D complexity hampers the utilization of critical applications over fading channels, such as incident wave angle estimation.} The incident wave angle, also known as the angle of incidence, is the joint azimuth and elevation angles between the direction of the propagation and the normal to the surface \textcolor{black}{of the antenna \cite{chen2010introduction}.}

Determining the incident wave angle is valuable for beamforming, directional antenna placement, localization, interference mitigation, network planning, as well as smart antenna systems. It enables improved signal quality, enhanced system capacity, reduced interference, and more efficient use of wireless resources. Contemporary methods for incident wave angle estimation consist of \textcolor{black}{the} use \textcolor{black}{of} either pilots on transmission \cite{ullah2021power}, or eigendecomposition of the received signal to subspace \cite{gentilho2020direction}. However, while the use of pilot signals reduces the maximum achievable rate, \textcolor{black}{the} eigendecomposition of the received signal becomes less effective due to the uncorrelated signal and known the number of path assumptions. Nevertheless, there are methods without requiring neither eigendecomposition nor additional pilots. Received signal strength (RSS) \textcolor{black}{has} been adopted in numerous studies for incident wave angle estimation, and it has been shown to be fairly successful \cite{malajner2011angle}.

\begin{table}[t]
    \centering
    \caption{\textcolor{black}{List of Parameters}}
      \resizebox{\linewidth}{!}{  
    \taburulecolor{black}
    \begin{tabu}{|l|p{0.45\textwidth}|}
        \hline
        \textbf{Parameter} & \textbf{Description} \\
        \hline
        $\textbf{A}_T^r, \textbf{A}_R^r$ & Tx and Rx position in rectangular coord. \\
        $\textbf{U}_T^r, \textbf{U}_R^r$ & Tx and Rx rotation in rectangular coord. \\
        $\textbf{U}_T^s, \textbf{U}_R^s$ & Tx and Rx rotation in spherical coord. \\
        $\tilde{\textbf{U}}_T^a, \tilde{\textbf{U}}_R^a$ & Oriented Tx and Rx into a state of ``a" \\
        $\theta_R^r, \theta_R^p, \psi_R$ & Roll, pitch and yaw for initial position \\
        $\phi^{\psi, \theta}_R$ & Incident wave angle with $\theta$ elevation and $\psi$ azimuth \\
        $\phi^{\theta}_R$ & Elevation component of the incident wave angle \\
        $\phi^{\psi}_R$ & Azimuth component of the incident wave angle \\
        $\tilde{\phi}^{\theta}_R$ & Estimated elevation component of incident wave angle \\
        $\tilde{\phi}^{\psi}_R$ & Estimated azimuth component of incident wave angle \\
        $F_T^E, F_R^E$ & $E-$plane antenna pattern for Tx and Rx \\
        $F_T^H, F_R^H$ & $H-$plane antenna pattern for Tx and Rx \\
        $\zeta$ & Antenna non-ideality factor \\
        $D(\psi,\theta)$ & Directivity of an antenna for $\phi^{\psi, \theta}_R$ \\
        $G(\psi,\theta)$ & Gain of an antenna for $\phi^{\psi, \theta}_R$ \\
        $h_{\ell,k}^\phi$ & $\ell-$th tap on $k-$th direction channel tap for $\phi^{\psi, \theta}_R$ \\
        $\kappa$ & Rician $K-$ factor \\
        $P_T$ & Transmitted signal strength \\
        $P_R$ & Received signal strength \\
        $p_\omega$ & Incident wave vector \\
        $p_A$ & Rx antenna vector \\
        $\mathcal{P}$ & Polarization loss factor \\
        $\hat{\theta}, \hat{\psi}$ & Misaligned $\theta$ and $\phi$ \\
        $\delta_{T1}, \delta_{T2}$ & Misalignment of Tx in $\theta$ and $\psi$ \\
        $\delta_{R1}, \delta_{R2}$ & Misalignment of Rx in $\theta$ and $\psi$ \\
        MAML($K$,$N$) & $N-$ class $K-$ shot MAML \\
        $\mathcal{Q}$ & Meta-model parameter \\
        $\hat{\mathcal{Q}}$ & Optimized meta-model parameter \\
        $\mathcal{Q}_{P_R}$ & Meta-model parameter for RSS \\
        $\mathcal{Q}^\circ$ & Meta-model parameter initialization \\
        $\phi^{\psi,\theta}_{\mathcal{Q}_{P_R}}$ & Estimated $\phi^{\psi,\theta}$ using meta-objective of ${\mathcal{Q}_{P_R}}$ \\
        $f_{\mathcal{Q}_{P_R}}$ & Meta-model classifier \\
        $\nu_{({\Phi_{R}^{\Psi,\Theta}})}$ & Meta-model classifier result \\
        $\epsilon$ & Meta model error tolerance \\
        $\alpha_{\tilde{\phi}_R}$ & Meta model confidence interval \\
        $\alpha$ & Meta-inner learning rate \\
        $\beta$ & Meta-step size \\
        \hline
    \end{tabu}
    }   
    \label{tab:parameters}
\end{table}
\begin{table*}[]
\caption{\textcolor{black}{Comparison of similar studies}}
\resizebox{\textwidth}{!}{%
 \taburulecolor{black}
\begin{tabu}{l|l|l|l|l|l|l|}
\cline{2-7}
 &
  \textbf{Antenna Modeling} &
  \textbf{Stochastic Channel Modeling} &
  \textbf{UAV Positoning} &
  \textbf{UAV Posture} &
  \textbf{Proposed Solution} &
  \textbf{Main Target Approach} \\ \hline
\multicolumn{1}{|l|}{\textbf{{}\cite{hua2023channel}{}}} &
  MIMO / Ideal &
  \begin{tabular}[c]{@{}l@{}}Air-to-Ground \\ Time-variant Rician Fading Channel \\ Free Space Path Loss\end{tabular} &
  Global Cartesian &
  \begin{tabular}[c]{@{}l@{}}Euler ZYX \\ (Roll, Pitch and Yaw)\end{tabular} &
  \begin{tabular}[c]{@{}l@{}}Model-based Neural Network \\ (Closed-form expression)\end{tabular} &
  \begin{tabular}[c]{@{}l@{}}Spatial Temporal Analysis \\ (Unknown measurement and dataset)\end{tabular} \\ \hline
\multicolumn{1}{|l|}{\textbf{{}\cite{mao20213d}{}}} &
  MIMO / Ideal &
  \begin{tabular}[c]{@{}l@{}}Air-to-Air\\ Time-variant Tapped Delay Line Channel\\ No Path Loss\end{tabular} &
  Local Cartesian &
  \begin{tabular}[c]{@{}l@{}}None\\  (No Posture Design)\end{tabular} &
  Time-variant scattering clusters &
  Spatial Temporal Analysis \\ \hline
\multicolumn{1}{|l|}{\textbf{{}\cite{zhao2023integrated}{}}} &
  MIMO / Ideal &
  \begin{tabular}[c]{@{}l@{}}Air-to-Ground \\ Time-invariant Physical Channel \\ No Path Loss\end{tabular} &
  Global Cartesian &
  \begin{tabular}[c]{@{}l@{}}Euler ZYX \\ (Roll, Pitch and Yaw)\end{tabular} &
  \begin{tabular}[c]{@{}l@{}}Channel Estimation and Sensor Fusion\\ with Integrated Sensing and Communication\end{tabular} &
  \begin{tabular}[c]{@{}l@{}}Impact of Attitude Variation\\ on Physical Channel\end{tabular} \\ \hline
\multicolumn{1}{|l|}{\textbf{Ours}} &
  \begin{tabular}[c]{@{}l@{}}SISO /\\ Ideal \& Non-Ideal\end{tabular} &
  \begin{tabular}[c]{@{}l@{}}Air-to-Air \\ Time-Invariant Rician Fading Channel \\ Log-normal Path Loss\end{tabular} &
  Global Cartesian &
  \begin{tabular}[c]{@{}l@{}}Euler ZYX \\ (Roll, Pitch and Yaw)\end{tabular} &
  \begin{tabular}[c]{@{}l@{}}Model Agnostic Meta Learning \\ (Open-form expression)\end{tabular} &
  Incident Wave Angle Analysis\\ \hline
\end{tabu}
}
\vspace{0.5cm}
\end{table*}
Note that the studies \cite{miranda2018enhanced,cheng2022optimal,zhang2022efficient} exploit the RSS in incident wave angle estimation \textcolor{black}{to} have the ideal antenna with stationary nodes assumptions. It is clear that the utilization of RSS for incident wave angle estimation in dynamic aerial nodes under antenna defects is not an easy task. Not only \textcolor{black}{does} the antenna radiation field \textcolor{black}{change} with each rotation, \textcolor{black}{but} antenna defects \textcolor{black}{disperse} the field characteristics of the antennas in aerial nodes. In addition, the transmitter aerial node also suffers from this challenge as well. In this perspective, autonomous or remotely operated unmanned aerial vehicles (UAVs) are under examination as aerial nodes that serve within an A2A network.

As a solution to this complexity challenge, artificial intelligence (AI) approaches can be adopted to mitigate incident wave angle estimation errors. Neural network based AI algorithms has the capability to optimize non-convex problems, and supervised learning can be deployed to map each RSS value to corresponding incident wave angles \cite{feng2022survey}. Convolutional neural networks (CNN) have been \textcolor{black}{used} in many aspects of wireless communication for estimation, detection, and data compressing \cite{wu2019cnn,luo2020machine}. Multiple studies show that channel adaptation of CNN is robust, although it fails in generalization for multi-task problems \cite{feng2021generalization,swiderski2022random}. \textcolor{black}{Alternatively, densely connected neural networks (DenseNet) \cite{huang2017densely} with a large number of parameters are formidable tools for learning complex patterns yet, not immune to the issue of out-of-sample distribution and struggle with generalization and rapid adaptation. Therefore, reliance on memorization and susceptibility to overfitting can limit their effectiveness in facing with unseen scenarios.}This leads to the search for new AI techniques that can create a generic AI model with limited data. \textcolor{black}{Therefore}, the \textcolor{black}{RSS-aided} incident wave angle estimation can be possible for any channel, antenna, and UAV conditions in A2A networks. 

There are contemporary solutions that can avoid model overfitting when using the limited amount of data provided. Few-shot learning technique as meta-learning is a learning-to-learn (L2L) algorithm that enables rapid convergence using \textcolor{black}{a} limited amount of samples \cite{vanschoren2019meta}. Meta-learning's practical use \textcolor{black}{was} developed in 2016 as an imitation of \textcolor{black}{the} human mind that \textcolor{black}{learns} things it has not encountered before \cite{andrychowicz2016learning}. Forming a generic predictive model is possible with model-agnostic meta-learning (MAML) that can divide the incident wave angle estimation problem into several tasks \cite{finn2017model}. \textcolor{black}{By this way}, an unknown task can be learned by the meta-optimization across the tasks by stochastic gradient descent (SGD). Based on these, the generated black box model can be evaluated by any independent model for the incident wave angle estimation task. Guided by this concept, MAML offers both flexibility and interpretability for the RSS mapping into the incident wave angles.

\subsection{Literature Review}

A directly related study \cite{9836318} examines the estimation of the direction of angle 
\textcolor{black}{in array antennas with linear polarization}. The importance of radiation patterns in aerial nodes has been highlighted in \cite{salari2021unmanned}. The study \cite{9208760} examines the cross-polarization of various UAV antennas experimentally and highlights the UAV scattering effects on radiation efficiency. Another related study \cite{zhou2022spatial} proves that fuselage posture is a key criterion in UAV-to-vehicle (U2V) networks. \textcolor{black}{A recent experimental air-to-ground study \cite{maeng2023impact} shows that taking the UAV antenna radiation pattern into account increases localization accuracy.} \textcolor{black}{The multi sensor state estimation of UAVs is also an ongoing study. The studies \cite{du2020real} and \cite{gamagedara2021quadrotor} \textcolor{black}{utilize} sensor fusion to \textcolor{black}{UAVs'} state estimation by inertial measurement sensors and global positioning system (GPS) information. State estimation with radio communication is a fresh topic, recently investigated in researches such as \cite{zhao2023joint} and \cite{wang2023communication}.} 

In this regard, the studies [17-24] primarily focus on the impact of antennas on UAV networks. While antennas are known to significantly impact aerial networks, existing studies often have limited consideration of antenna imperfections. For instance, [17] uses multi-mode multi input multi output (MIMO) antennas for the direction of arrival estimation, but the simulation-demonstration disparity is notable. Similarly, [18] and [19] discuss antenna issues such as cross-polarization, and [20] emphasizes the influence of fuselage posture on spatial-temporal correlations in a Rician fading aerial channel.

In our study, we push a step further in A2A communication by considering antenna imperfections in angle estimation. Recognizing the crucial roles that both the channel and signal quality play, we delve into their impact on localization. Utilization of RSS is substantial in incident wave angle estimation techniques \cite{cheng2022optimal,bnilam2020rss}. One study \cite{tomic2018target} examines the target localization by RSS and incident wave angle information only. Nevertheless, the dynamics of the UAVs are not taken into consideration for these studies. Nevertheless, maneuvering, leading/following a trajectory, \textcolor{black}{and} wobbling are some common behaviors of all UAVs. Furthermore, it has been shown that the banana distribution of a hovering UAV motion is a special Gaussian distribution as well \cite{long2013banana}. A recent study \cite{amodu2023thz} highlights the fluctuation degradation and mobility angle impact over the link capacity for terahertz (THz) communication in A2A networks.

The use of few-shot learning was found to be very limited in wave characterization. \textcolor{black}{ Furthermore, the exploration of A2A network analysis in the context of UAV antenna imperfection remains a relatively unexplored and unaddressed domain within the scientific community. As a result, the potential effects of incident wave angle estimation in realistic scenarios have not been thoroughly investigated, leaving this aspect largely uncharted. Additionally, the application of AI to tackle and alleviate such challenges remains largely undiscovered, providing an opportunity for further research and innovative solutions in this field.} In the sense of meta-learning usage in related topics, a recent study utilizes meta-learning in beam prediction \textcolor{black}{problems} for multiple input-single output (MISO) channel \cite{yang2022meta}. In \textcolor{black}{a} similar context, \cite{xia2021meta} exploits the meta-learning in maximization of the weighted sum rate problem for beamforming. One another study \cite{9745789} uses meta-reinforcement learning to maintain user positions for virtual reality (VR) networks. \textcolor{black}{One study makes use of the few-shot learning for multimodal target detection \cite{khoshboresh2023multimodal}. In \cite{wu2022application}, a first order model agnostic meta-learning (FOMAML) and almost no inner loop (ANIL) techniques have been used as autoencoder for \textcolor{black}{RIS-assisted} air-to-ground UAV networks. Evolution strategies for meta-learning variants such as reptile and meta-reinforcement are discussed in \cite{majid2023deep}. An overview of the communication techniques with the meta-learning technique is examined in \cite{chen2023learning}. In robotics, meta-learning is proposed for trajectory optimization in UAVs under challenging conditions \cite{yel2021meta}. }Apart from \textcolor{black}{the} L2L field, usage of machine learning in beamforming technique is widespread, as in the contributions \cite{b1,b2,b3}.

Two most related studies are compared in Table II with this research in terms of the methodology and the aim. Regarding the orientation formation, this study and \cite{hua2023channel} use the common terms of ``elevation" and ``azimuth'' which only can be described with ``$Z-Y-X$" Euler representation in UAV posture definitions. In \cite{hua2023channel}, a machine learning algorithm utilizes channel state information (CSI) such as path delays, path loss, shadow fading and Doppler shift. On the other hand, this study keeps a simple time invariant case to go deep further to generate a consistent dataset of RSS with a non-stationary quasi-static channel model, in order to estimate incident wave angles.

The use of RSS offers two distinct advantages. Firstly, being a non-complex value, it places less demand on the neural network, streamlining computational requirements. Secondly, the receiver UAV benefits from not having to engage in post-processing tasks, such as signal decryption which preserves privacy and saves from the over-the-air computation. In \cite{mao20213d}, the authors aims to introduce a geometry-based stochastic approach involving spatial dynamic clustering among UAVs. While the cluster evolves over time, the study does not account for UAV postures. Consequently, the \textcolor{black}{maneuvers} and rotations of the UAVs are overlooked, and correlations rely solely on the arbitrary trajectory each UAV follows. However, \textcolor{black}{our} study, in alignment with findings from \cite{hua2023channel}, shows that UAV posture is non-stationary and any rotation of the UAV body significantly influences signal quality.

Briefly, the existing literature lacks a comprehensive 3D channel model that incorporates antenna imperfections for dynamic A2A networks, as well as a solution for estimating the incident wave angle. The few available studies with meta-learning for incident wave angle estimation are limited in scope, as they do not explore both elevation and azimuth incident wave angles, and do not address the dynamic UAV antenna responses. In light of this research gap, this study aims to provide a thorough definition and a comprehensive solution for this unexplored area within A2A networks.

\textcolor{black}{Based on the findings, this research makes the following contributions:}
\begin{itemize}
    \item Modeling a generic 3D channel model under antenna imperfections with polarization loss, antenna mismatches and antenna misalignment for A2A networks. Impacts of position and orientation are included. 
    \item \textcolor{black}{Proposing few shot meta-learning approach for estimating incident wave angles within an unknown channel by adopting a multitask learning perspective with limited RSS information.} Meta-learning steps for MAML have been explicitly derived for the proposed incident wave angle estimation method. \textcolor{black}{A custom dataset for the task optimization is created.}
    \item \textcolor{black}{Adopting} a lower computational cost FOMAML technique in beam classification as an alternate to MAML and evaluating the impact of less gradient information on the incident wave angle estimation.
    \item Deriving a bound for the probability of confidence (PoC) of the generated AI models by Hoeffding's inequality to justify their validity and reveal the requirement of meta-test dataset size for \textcolor{black}{the} desired success rate.
    \item Evaluating the generated models with the proposed dataset and demonstrating comparisons with each other, as well as \textcolor{black}{the CNN model of MAML and more sophisticated $\mathrm{DenseNet}$ network}. 
\end{itemize}

\subsection{Organization and Notations}
The rest of the paper is organized as follows. In Section \ref{sec2} the system model with dynamic 3D-UAV integrated hardware and channel model has been introduced. In Section \ref{sec33}, the challenges of the system and drawbacks within the contemporary solutions have been discussed. Next, Section \ref{sec4} establishes a collective task learning model with meta-learning in fast adaptation of any fading channel models. The numerical results have been discussed in Section \ref{secnum} and the research directions that proposed model lead are listed in the Section \ref{secopen}. Finally, the conclusion of this study have been briefed in Section \ref{sec7con}. 

\textit{Notation}: While bold letters have been denoted as matrices, $|| \cdot ||_2$ indicates the $L^2$ norm and $\mathbb{CN}(0,\sigma^2)$ is the complex Gaussian distribution with zero mean and $\sigma^2$ variance. $\mathbb{U}(a,b)$ notates the uniform distribution with the lowest value $a$ and highest value $b$. $\mathbb{E}[\cdot]$ operator is the expectation, $p(\mathcal{X})$ is the probability density function of a random variable $\mathcal{X}$. While $\textbf{I}^{n\times n}$ and $\textbf{0}^{n\times n}$ are the identity matrix and zero matrix with order $n$ respectively, $\Pr(\cdot)$ is the probability function, $\exp(\cdot)$ is the exponential function, $\mathbbm{1}$ is the indicator function and $\mathcal{X}^\intercal$ is the transpose of $\mathcal{X}$ and $\textbf{proj}_b a$ indicates the projection of $a$ on the plane $b$.
\label{sec2}

\section{System Model}
\label{sec2}
\begin{figure}[t] 
    \includegraphics[width=0.5\textwidth]{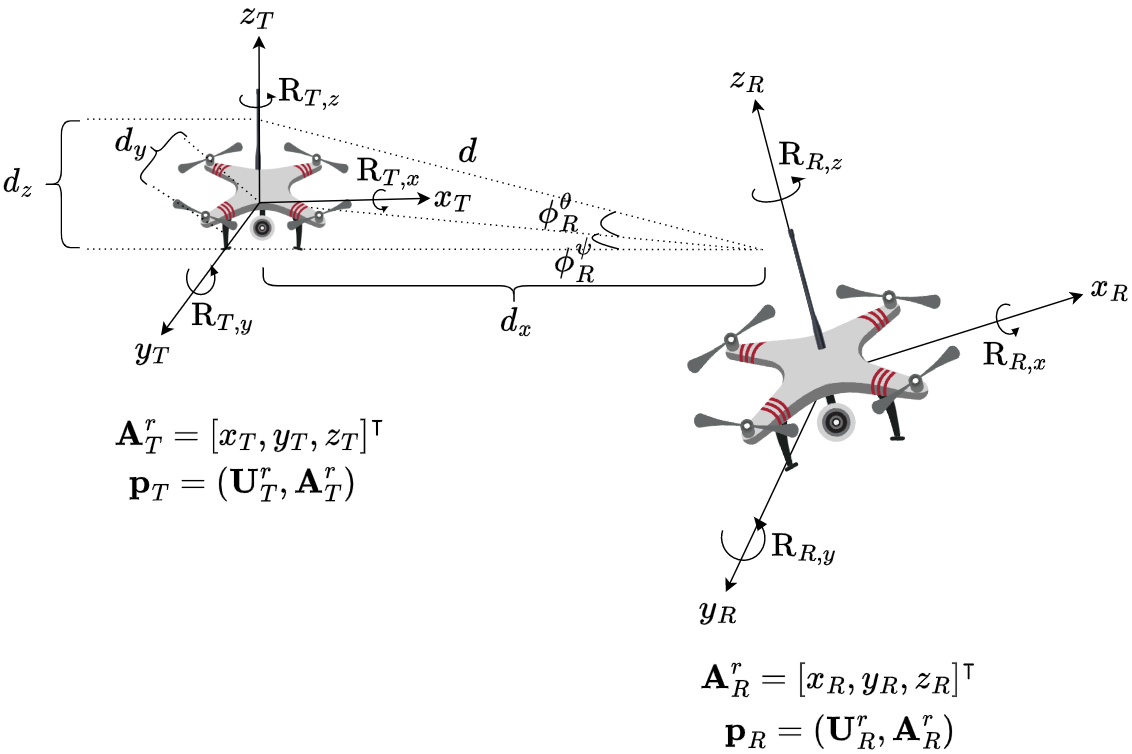}
    \caption{An ideal dynamic node-to-node A2A propagation model with on-board short dipole antennas in A2A networks.}
    \label{drones}
\end{figure}

A fair and realistic A2A propagation between 3D positioned UAV nodes can be shown on the Cartesian coordinates to analyze the link budget and radio frequency (RF) propagation. A UAV node that can be the transmitter (Tx) or the receiver (Rx) with 3D rectangular position vectors $\textbf{A}^r_{T} = [x_T, y_T, z_T]^\intercal$, $\textbf{A}^r_{R} = [x_R, y_R, z_R]^\intercal$ and rotation vectors $\textbf{U}^r_T = [\hat{e}_{x_T}, \hat{e}_{y_T}, \hat{e}_{z_T}]^\intercal, \textbf{U}^r_R = [\hat{e}_{x_R}, \hat{e}_{y_R}, \hat{e}_{z_R}]^\intercal $ such that $\theta_R^r$ and $\theta_R^p$ are the roll and pitch angle that corresponds to the elevation angle for the UAV antenna and $\psi_R$ is the yaw angle where it corresponds to the azimuth angle\footnote{Note that the azimuth and elevation angles cannot represent the 3D physical orientation relative to a reference point in the coordinate system. In this case, the described initial position of the antenna enables us to make this definition. Since the $\theta_R^r$ rotation (heading) has no impact upon the radiation due to the omnidirectivity, the elevation corresponds to $\theta_R$ and yaw corresponds to $\psi_R$ azimuth angle for the given initial position at the receiver.} for the UAV antenna, with the assumption that the UAV hovers over $xy$- plane with its antenna along $z$- axis. First, the orientation with 6-degrees of freedom over \textcolor{black}{the} principle axis has been analyzed on the line-of-sight (LoS) path in the link budget. The impact of the hovering and rotating of \textcolor{black}{UAVs} on a non-stationary quasi-static channel will be discussed. Note that body-fixed reference frame (BRF) and global reference frame (GRF) have been used to analyze the state of the UAV nodes.

\subsection{Dynamic Antenna Model} 
\begin{figure}[] 
  \includegraphics[width=0.5\textwidth]{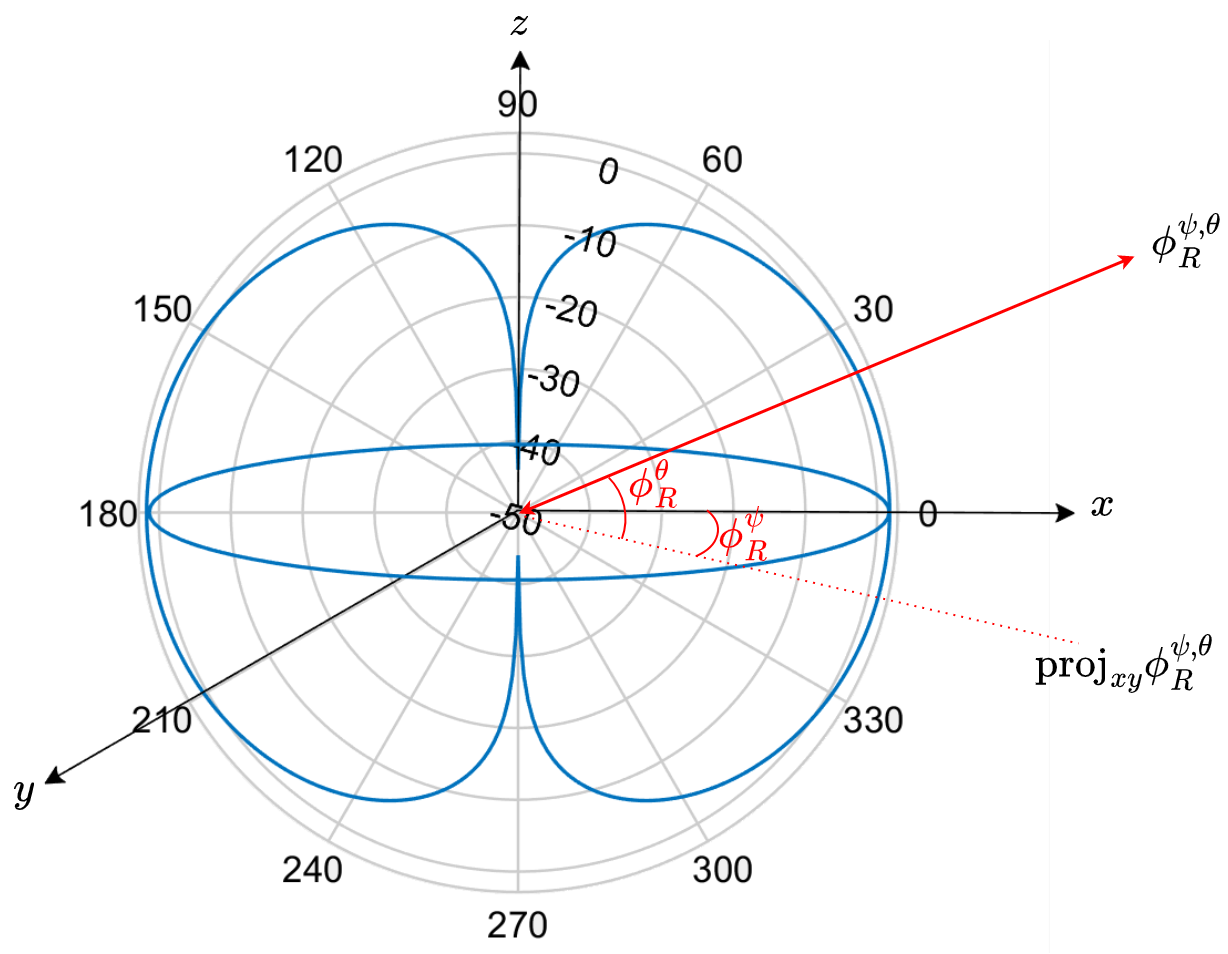}
    \caption{3D representation of incident wave upon an ideal $\textbf{U}_R^r$ antenna radiation.}
    \label{incidentwave}
\end{figure}

\label{1
}
An embedded probe dipole antenna on a UAV has been taken into consideration for radio transmission and reception. Therefore, it has been assumed that the center-fed omnidirectional short dipole antenna has the same rotation vector with the $\textbf{A}_R$, and the boresight is in the symmetry axis of the antenna. In order to study the polarization effect in depth, it was deemed appropriate to use a $L$ length cosine pattern short dipole with linear polarization. It should be highlighted that this study only concerns the far-field. \textcolor{black}{ A single input single output (SISO) configuration was used instead of a MIMO structure to more effectively exploit the benefits of the AI solution technique. The advantages of MIMO setups may not be as pronounced as in SISO for LoS dominant A2A systems. In a LoS scenario, spatial diversity can be diminished due to the presence of a direct path between the antennas. 
\textcolor{black}{Spatial multiplexing gains of MIMO} are typically more pronounced in environments with rich scattering and multipath, where signals take multiple paths with different delays and phases. Moreover, the separation, weight, and size of antennas are constrained by the physical limitations of the UAV body. However, in a more advanced A2A MIMO system, the array of options for determining wave angles can significantly increase.
}
For simplicity, 3D rotation of the Tx with $\textbf{A}^r_T =[x_T,y_T,z_T]^\intercal$ can be considered as immobile over a $xy-$ plane while the Rx is randomly spatially located in GRF with desired orientation vector $\textbf{U}_R^r$. Thereafter, the electromagnetic loss of the Rx with $\textbf{A}_R^r$ in $\textbf{U}_R^r$ will be derived such that the roll and pitch turns in Tx and/or Rx lead to gradual nulling and polarization loss i.e. $\theta_R=90\degree$ rotation around $x-$axis where the antenna polarization orthogonal to the incident wave angle $\phi_R^{\psi,\theta}$ illustrated in Figure \ref{incidentwave}. The spherical basis matrix for Tx ($\textbf{U}^s_T$) and Rx ($\textbf{U}^s_R$) is $\textbf{U}^s_{T,R}= [\hat{e}_R, \hat{e}_\psi, \hat{e}_\theta]^\intercal$ such that,
\begin{align}
    \hat{e}_R &= \cos{(\theta)}\cos{(\psi)}\hat{\textbf{i}} +             \cos{(\theta)}\sin{(\psi)}\hat{\textbf{j}}+\sin{(\theta)}\hat{\textbf{k}},\nonumber \\
    \hat{e}_\psi &= -\sin{(\psi)}\hat{\textbf{i}} + \cos{(\psi)}\hat{\textbf{j}},\\
    \hat{e}_\theta &= -\sin{(\theta)}\cos{(\psi)}\hat{\textbf{i}} - \sin{(\theta)}\sin{(\psi)}\hat{\textbf{j}} + \cos{(\theta)}\hat{\textbf{k}} \nonumber,
\end{align}
\textcolor{black}{which can be used for coordinate transformations. Here, $\hat{\textbf{i}},\hat{\textbf{j}}$ and $\hat{\textbf{k}}$ are} positive direction unit vectors. Furthermore, the UAV (thereby the antenna) rotation can be shown in $Z-Y-X$ Euler angle sequence 
\begin{equation}    
    \tilde{\textbf{U}}^a_{T,R}= \underbrace{\textrm{R}_z(\Theta_1) \textrm{R}_y(\Theta_2) \textrm{R}_x(\Theta_3)}_{\textbf{R}^{ab}} \textbf{U}^b_{T,R},
\end{equation}
where $\textbf{U}^b_{T,R}$ is the initial attitude of the node in BRF, $\tilde{\textbf{U}}^a_{T,R}$ is the orientation of rotated UAV node and $R_\gamma(\Theta)$ is the unimodular rotation matrix as $\textbf{R}:\mathbb{R}^{n\times n}\rightarrow\mathbb{R}^{n\times n}$ and $\textbf{R}_\gamma(\Theta)$ has been defined in Eq. (7) along with its components
where $\textrm{R}_\gamma^{-1}(\Theta) = \textrm{R}_\gamma(-\Theta)$ and the member of special orthogonal group with $SO(3)=\{\textbf{R}\in \mathbb{R}^{n\times n}|\textbf{R}^T\textbf{R}=1,\textrm{det}(\textbf{R})=1\}$. Rotation around the $z$-axis with positive $\Theta_1$ angle is the azimuth angle, rotation around the new $y$-axis with positive $\Theta_2$ angle is the pitch and lastly, rotation around \textcolor{black}{the} new $x$-axis with positive $\Theta_3$ angle is the roll definitions on the UAVs. \textcolor{black}{The parameters of interest included in UAVs state space defined as $\textbf{p}_{T,R}=(\textbf{U}_{T,R}^r,\textbf{A}_{T,R}^r)$ as in the Figure \ref{drones}. \textcolor{black}{
For the flight stabilization conventions, UAVs always kept stabilized before any
\textcolor{black}{maneuvering}. A UAV takes the path from a point $A$ to $C$ over a travelling point of $B$ (e.g. $A\rightarrow B \rightarrow C$) kept stabilized to initial pose orientation. The UAV takes the state of $a$ as
$\tilde{\textbf{U}}^a_{T,R}= {\textbf{R}^{ab}} \textbf{U}^b_{T,R}$, and then recovers to $\tilde{\textbf{U}}^b_{T,R}= {\textbf{R}^{ba}} \textbf{U}^a_{T,R}$. This completes the $A \rightarrow B$. Continuing on, the same logic repeats from point $B$ to $C$. In this path, UAV takes the state of $c$ for $B \rightarrow C$ path as $\tilde{\textbf{U}}^c_{T,R}= {\textbf{R}^{cb}} \textbf{U}^b_{T,R}$.} Note that it has been assumed that receiver UAVs have the self-orientation information via inertial proprioceptive sensors.} Based on these rotations, the radiation pattern of Rx with $\tilde{\textbf{U}}^a_{T,R}$ can be defined with \textcolor{black}{the} decomposition of the principal patterns of $E$- plane and $H$- plane components in GRF as following
\begin{equation}
    \textbf{F}_{T,R}(\psi,\theta,\zeta) = [F^E_{T,R} (\psi,\theta,\zeta), F^H_{T,R}(\psi,\theta,\zeta)]^\intercal.
    \label{impairk}
\end{equation}
The antenna follows additive $\zeta \sim \mathbb{N}(0,\sigma^2_\zeta)$ impairment model for sourced by fabrication and reflections that will be discussed in Section \ref{secimpa}. Showing the dynamic model of antenna position and orientation in Cartesian coordinates, A2A distance is defined as $d = ||\textbf{A}^r_T-\textbf{A}^r_R||_2$ and $\phi_R^\psi = \arctan\bigl({\frac{d_y}{d}}\bigl)$ and $\phi_R^\theta = \arctan\bigl({\frac{d_z}{d}}\bigl)$. Since the antenna pattern covers in $(0, 2\pi]$ azimuth field of view is equal in \textcolor{black}{an} ideal case, $\phi_R^\theta$ will be the main concern of this study. The probability of UAV's position is considered as uniform distribution with near-field constraint ($d_{\textrm{min}}$) and coverage limit ($d_{\textrm{max}}$) where 
\begin{equation}
    p_d(d)=\frac{1}{d_{\textrm{max}}-d_{\textrm{min}}}, \hspace{2mm}  [d_{\textrm{min}}<d<d_{\textrm{max}}].
\end{equation}
As a result, the distribution of $p_{\phi^\theta}(\phi^\theta|d)$ can be obtained with transformation density function as 
\begin{equation}
    \begin{split}      
    p_{\phi^\theta}(\phi^\theta|d)& = p_d\bigg(\frac{d_z}{\tan(\phi^\theta)}\bigg)\bigg|\frac{\partial d}{\partial \phi^\theta}\bigg|, \\
    &=p_d\bigg(\frac{d_z}{\tan(\phi^\theta)}\bigg)\big|d_z \csc^2(\phi^\theta)\big|,\\
    &=\frac{d_z}{d_{\textrm{max}}-d_{\textrm{min}}} \csc^2(\phi^\theta),
    \end{split}
    \label{pdff}
\end{equation}
where $\phi^\theta$ is denoted by $\Bigr\{\phi^\theta \in \mathbb{R}: \arctan{\Bigl(\frac{d_z}{d_{\textrm{min}}}\Bigl)} < \phi^\theta < \frac{\pi}{2}\Bigl\}$. With the linear primary polarization, the limitations in the gain ($G$) can be obtained with the knowledge of transmission $\phi_T^{\psi,\theta}$ and incident wave angles $\phi_R^{\psi,\theta}$.

\textbf{Lemma 1.} A short dipole antenna with $L<<\frac{\lambda}{2}$, the radiation intensity ($U$) is a function of \textcolor{black}{the} electric field $E_\theta$ in \textcolor{black}{far-field} radiation region and therefore the directivity is a function of $\phi^\theta_{T,R}$ as following\textit{ 
\begin{align}
    &D(\psi,\theta) =  4 \pi \frac{\max U(\psi,\theta)}{P_r} = 4 \pi \frac{\max U(\theta)}{P_r}, \nonumber \\
    &G(\psi,\theta) = D(\psi,\theta).
\end{align}
Proof. }See Appendix A.

\setcounter{equation}{6}
  \begin{figure*}
  	\begin{multline}	    
  	\label{eq:5}
            \textrm{R}_x(\Theta_3)=
            \begin{pmatrix}
            1 & 0 & 0\\
            0 & \cos(\Theta_3) & -\sin(\Theta_3) \\
            0 & \sin(\Theta_3) & \cos(\Theta_3)
            \end{pmatrix},            \textrm{R}_y(\Theta_2)=
            \begin{pmatrix}
            \cos(\Theta_2) & 0 & \sin(\Theta_2)\\
             0 & 1 & 0 \\
            -\sin(\Theta_2) & 0 & \cos(\Theta_2)
            \end{pmatrix},
            \textrm{R}_z(\Theta_1)=
            \begin{pmatrix}
            \cos(\Theta_1) & -\sin(\Theta_1) & 0\\
            \sin(\Theta_1) & \cos(\Theta_1) & 0 \\
            0 & 0 & 1
            \end{pmatrix}
            \\
  	\end{multline}
     	\noindent\rule{\textwidth}{.5pt}
  \end{figure*}
\setcounter{equation}{7}  

\begin{figure}[t] 
    \includegraphics[width=0.5\textwidth]{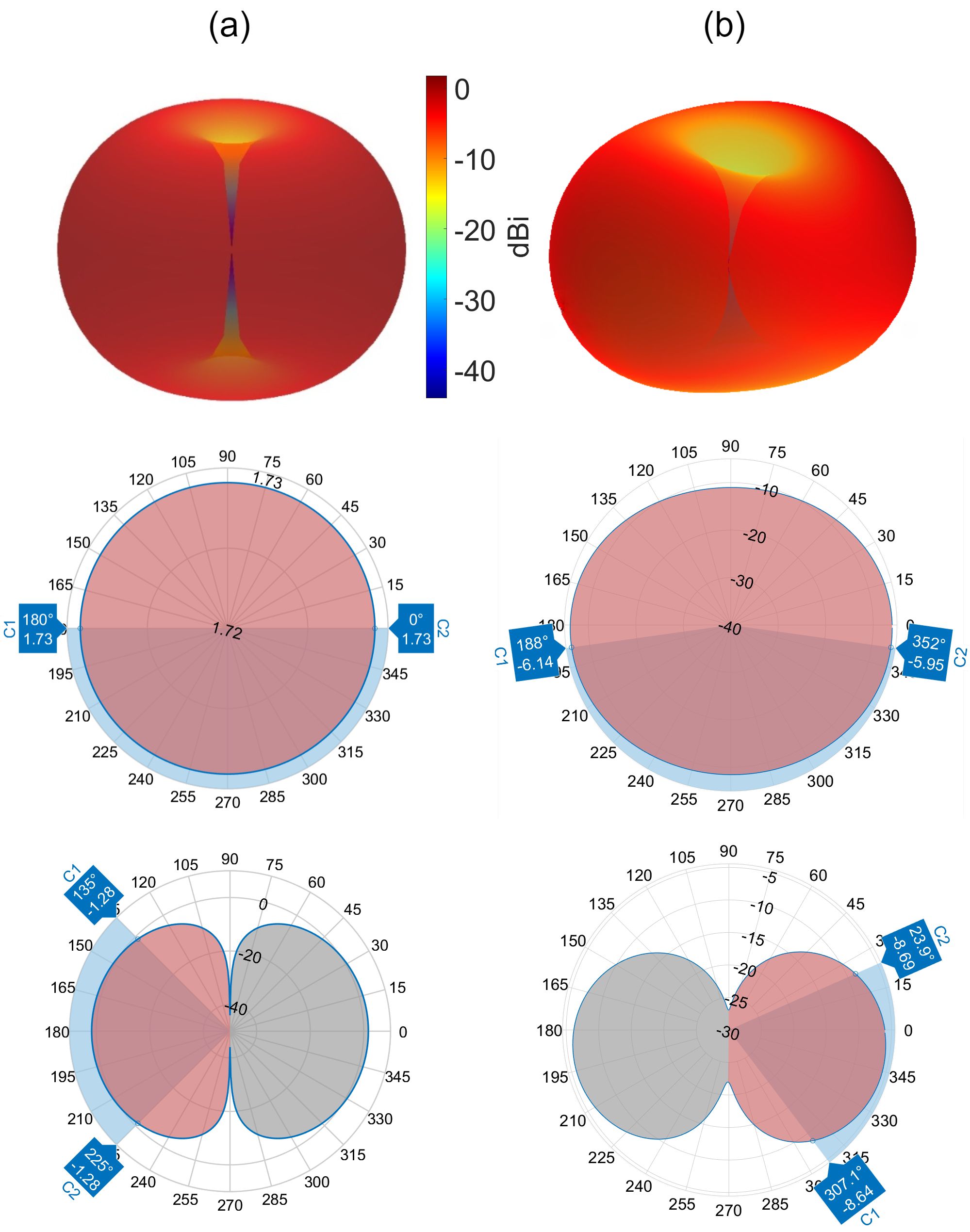}
    \caption{Impact of antenna imperfectivity to the radiation pattern. a) Ideal Antenna vs. b) Non-ideal antenna with $\zeta$ impairment.}
    \label{impairment}
\end{figure}
\subsection{Channel Model}

An accurate estimation of incident wave angle offers the feasibility \textcolor{black}{of adapting} the angle varying channel. A SISO configured A2A link is considered for this study. Both uplink and downlink transmissions are taken into account during the 3D channel forming. Channel vector has \textcolor{black}{a} dependency on 
the $\phi_R$ denoted as $\textbf{h}_k^\phi \in \mathbb{C}^{1\times N}$ where $N$ is the number of multipath components (MPC). \textcolor{black}{The system model to be described holds true for bidirectional propagation, as both the channel and the antenna are reciprocal. In this respect, the communication direction is denoted by a transmission index $k = \{1,2\}$.} The received signal $y_k^\phi$ can be shown as \textcolor{black}{follows}
\begin{equation}
    y_k^\phi = h_k^\phi x_k + n_k^\phi,
\end{equation}
where $x_k$ is the transmitted signal, $\textbf{n}_k^{\phi} \sim \mathbb{CN}(0,\sigma^2)$ is the complex additive white Gaussian noise (AWGN). With the assumption of having \textcolor{black}{an} LoS link, \textcolor{black}{the frequency selective Rician fading channel with \textcolor{black}{MPC} under \textcolor{black}{an} LoS link has been considered for this study.} \textcolor{black}{ Due to \textcolor{black}{the} predominant LoS path, the Rician fading channel model is used which also takes the scattering and multipaths account.} Therefore, a discretized $L$-tap baseband model with bandwidth $W$ and $m/W$ sampling, baseband channel $g_{\ell,k}$ can be shown as following \textcolor{black}{\cite{tse2005fundamentals}}:
\begin{equation} \label{eq1}
    g_{\ell,k}[m]=\sum_{i} a_i (m/W) e^{-j2\pi f_c \tau_i (m/W)} \sinc[\mathcal{\ell}-\tau_i (m/W)W],
\end{equation}
where $f_c$ is the centre frequency, the gains of $\ell-$th tap is $a_i$ and propagation delay is $\tau_i$ follows uncorrelated wide-sense stationary scattering within single tap.
Thence, the distribution of cumulated independent filter taps forms a Rician fading with $\kappa$ -factor can be written as
\begin{equation}
g_{\ell,k} [m]= \sqrt{\frac{\kappa}{\kappa+1}}\sigma_\ell e^{j\xi} + \sqrt{\frac{1}{\kappa +1}}\mathbb{CN}(0,\sigma^2_\ell),  
\end{equation}
where the first term includes the incident wave characteristics ($\xi$) of the LoS component while the second term is the circular symmetric cumulated diffuse components, independent from the first LoS component, therefore the $\phi^{\psi,\theta}$. Expanding the $\xi$, channel of \textcolor{black}{an} LoS incident wave angle under dynamic antenna configuration with $\mathbb{E}[|g_{\ell,k}|^2]=1$ can be described with as following
\begin{multline}
g_{\ell,k}^{\phi} [m] = \sqrt{\frac{\kappa}{\kappa+1}}\sigma_\ell \exp\{j2\pi f_D \cos{(\phi_R^{\psi(\cdot),\theta})}
+\varphi\}\\ 
+ \sqrt{\frac{1}{\kappa +1}}\mathbb{CN}(0,\sigma^2_\ell), 
\end{multline}
where $f_D$ notates the maximum Doppler frequency, \textcolor{black}{$\phi_R^{\psi(\cdot),\theta}$ is the incident wave angle with $\theta$ and arbitrary azimuth angle $\psi$}, and $\varphi$ is the phase. The UAV azimuth rotations around $z$-axis impact for $P_R$ can be neglected on GRF such that $F_{R}(\psi_R^i,\theta_R^{(a)}) = F_{R}(\psi_R^j,\theta_R^{(b)}), \forall{i,j}$ if and only if $\theta_R^{(a)} = \theta_R^{(b)}$. Therefore, the fading channel by using the predefined geometry for A2A becomes,
\begin{multline}
g_{\ell,k}^{\phi} [m] = \sqrt{\frac{\kappa}{\kappa+1}}\sigma_\ell \exp\Bigl\{j2\pi f_D \cos{\Bigl(\arctan\Bigl({\frac{d_z}{d}}\Bigl)\Bigl)}
+\varphi\Bigl\}\\ 
+ \sqrt{\frac{1}{\kappa +1}}\mathbb{CN}(0,\sigma^2_\ell). 
\end{multline}
For the large scale fading model, outdoor log-distance path loss is defined as 
\begin{equation}
    \textrm{PL (dB)} = \textrm{{PL}}_0 + 10 \eta \log_{10} \Bigl(\frac{d}{d_0}\Bigl) + X_g,
\end{equation}
where $d_0$ is the reference distance, $\textrm{{PL}}_0$ is the free-space path loss in dB, $d$ is the length of path, $\eta$ is the path loss exponent and $X_g \sim \mathbb{CN}(0,\sigma^2_p)$ is the fading loss by the described channel model. The corresponding large-scale average channel gain for the $\ell$-th tap is $c_{\ell,k}[m]=d^{-\eta} 10^{\frac{-\text{PL}_0}{10}}$, independent of the $\phi^{\psi,\theta}$. Thus, the accumulated channel for two-way channel with $\phi^{\psi,\theta}$ incident wave is following
\begin{equation}
    h_{\ell,k}^\phi [m] = \sqrt{c_{\ell,k}[m]}g_{\ell,k}^\phi [m]. 
\end{equation}
Simply, RSS is defined as $P_R = P_T - \textrm{PL}$ at the Rx UAV. The SNR for such a two-way A2A with $\phi^{\psi,\theta}$ link model without interference can be defined as following,
\begin{equation}
    \textrm{SNR}^\phi_{\ell,k} = \frac{P_T |h_{\ell,k}^\phi |^2 G(\psi, \theta)}{\sigma^2},
\end{equation}
where $|h_{\ell,k}|^2$ is the channel gain at the corresponding node.
\section{\textcolor{black}{An Overview of the Impairments and Estimation Problems}}
\label{sec33}
As one of the main \textcolor{black}{focuses} for localization application requirements, incident wave angle estimation by using time-of-flight (ToF) techniques such as time difference of arrival (TDoA) between the aerial nodes is quite affordable to implement. However, extracting the elevation and azimuth angles out of the incident wave angle is also an estimation problem to solve that has been discussed in \cite{al2017two}, \cite{wu2019multiemitter}. Apart from the estimation errors, the proposed methods in those studies are defined under some constraints (e.g. randomness of the positions in the array elements, only viable in a limited span). The root of this fundamental problem which is the propagation direction of the incident wave of \textcolor{black}{the} UAV node can be found by the RSS. \textcolor{black}{With the conventional methods \cite{miranda2018enhanced}, minimum differential RSS ($\Delta P_R$) leads to the incident wave angle for the reference plane. Note that ${\textbf{P}_R} =[P_R^{(1)},...,P_R^{(K)}]^\intercal$ and $\Delta P_R = P_R^{(i)} - P_R^{(j)}$. For the vertically polarized short dipole antenna, minimized $\Delta P_R$ with the corresponding maximum power $\Lambda(\phi)=|F^E_{R} (\psi,\theta,\zeta)|^2$ that leads to the $\hat{\phi}_R$ solution modeled as the following, }
\begin{equation}
    C(\phi) = \frac{1}{||\mathbf{\Lambda}(\phi) - \mathbf{\Delta} P_R||_2^2}.
\end{equation}
\textcolor{black}{where $\Lambda(\phi) \subset \mathbf{\Lambda}(\phi)$ and $\Delta P_R \subset \mathbf{\Delta} P_R$. As a result, the estimated $\hat{\phi}_R$ is following}
\begin{equation}
    \hat{\phi}_R = \argmax_{\phi_R \in \Phi_R} C(\phi).
\end{equation}
With this approach, Tx orientation $\tilde{\textbf{U}}^b_T$ is not required to be known by the Rx, simply can be estimated with \textcolor{black}{the} RSS maximization algorithm for ideal Tx and Rx antennas. 

There are many major challenges \textcolor{black}{to} having an accurate $\hat{\phi}_R$ for A2A networks. Among the ToF techniques, one essential need for TDoA is the \textit{synchronization} of sensor clocks in universal time during the localization process. Another requirement is the \textit{zero misalignment} between the aerial nodes, which is difficult to conceptualize in A2A networks. Conventional AI with data driven methods is not a feasible solution since $p_{\phi^\theta}(\phi^\theta|d)$ for each trajectory sets in $\mathcal{S}$ with a different distribution which only can be defined as a new task. \textcolor{black}{As a result, each solution offered therefore brings with it another problem and/or constraint.} We search for a way to approach the most generic solution with \textcolor{black}{the} minimum amount of dependency in A2A networks that \textcolor{black}{contain} hardware challenges \textcolor{black}{from a} 3D channel perspective. Then we evaluate the generalized model for incident wave angle estimation in 3D channel models exposed by the antenna defects and imperfections. 


Some main concerns related \textcolor{black}{to} antenna imperfections namely polarization loss due to antenna mismatches, antenna misalignment and RF impairments due to the reflections and fabrication errors are described in \textcolor{black}{the} following subsections. 
\textcolor{black}{``A general comment can be made that the results of the impairment models are influenced by aggregated independent non-idealities, leading to propagated errors."} \textcolor{black}{
Accordingly, the impact of the antenna imperfections in investigated with their direct results on the system model. Firstly, the orientations of UAVs result in polarization loss, only if the orthogonality between the antenna and incident wave angle is broken down. Secondly,
the positioning of the UAVs may result within 2 different scenarios. 1) UAVs are aligned but not orthogonally placed on the space (due to a path trajectory or mobility model) intentionally and 2) UAVs are not aligned to each other due to \textcolor{black}{maneuvering} and wobbling unintentionally on any placement. On both scenarios, the misalignment results \textcolor{black}{in} the radiation loss on \textcolor{black}{the} received signal. Therefore, we kept our investigation on polarization, misalignment and fabrication losses which ultimately covers any loss scenarios.}
\subsection{Polarization Efficiency} 
The polarization matching in LoS links in A2A communication becomes a significant challenge considering the UAV rotations. Only vertical or horizontally polarized antenna transmissions with rotation on the orthogonal axis will lead to polarization loss. Considering the impact on the $P_R$, this loss determines the throughout impact of the mismatch, without requiring to deep through analysis \cite{9836318} of field polarization for each plane. Antenna polarization impact on both $\textbf{U}_T^r$ and $\textbf{U}_R^r$ can be investigated on the same coordinate system as the polarization vector of the wave. The polarization loss factor (PLF) can be defined as 
\begin{equation}
    \mathcal{P}=|p_\omega p_A|^2, 
\end{equation}
where $p_\omega \in \mathbb{R}$ is the incident wave polarization, $p_A \in \mathbb{R}$ is the receiver antenna polarization vector and $0\leq \mathcal{P}\leq 1$. On the 3D perspective with $\hat{\theta}_R$ and $\hat{\psi}_R$ antenna misalignment components, the PLF for linearly polarized antenna components can be shown as 
\begin{equation}
    \begin{cases}
    \mathcal{P}_\psi = \textbf{U}_{T,R}^{r,\textrm{PLF}({\psi})} = \cos^2{\hat{\psi}_R},\\
    \mathcal{P}_\theta = \textbf{U}_{T,R}^{r,\textrm{PLF}({\theta})} = \cos^2{\hat{\theta}_R}.
    \end{cases}
\end{equation}
As a result, the weakened received signal by the PLF denoted as following
\begin{equation}
P_R^\dagger = \mathcal{P_\psi} \mathcal{P_\theta}P_R.
\end{equation}
For example, it is more than enough \textcolor{black}{to have} one orthogonal axis rotation with respect to $\textbf{U}_R^r$ to maximize the PLF. 
In addition, these impairments have an impact on each other. \textcolor{black}{
`` On the ideal case of polarization loss factor (PLF), heading (yaw axis) of the short dipole UAV antennas does not change the orthogonality between two UAV nodes since $\textbf{F}_{T,R}(\psi,\theta) = \textbf{F}_{T,R}(\psi+\delta \psi,\theta)$. However, this is not the case on non-ideal antenna, where $\textbf{F}_{T,R}(\psi,\theta,\zeta(\psi,\theta)) \neq \textbf{F}_{T,R}(\psi+\delta \psi,\theta,\zeta(\psi+\delta \psi,\theta))$. With an unbiased impairment assumption, the antenna omnidirectivity allows the yaw independency for the large amount RSS instance \cite{gallager2013stochastic} if and only if $\delta \sim \mathbb{N}(0,\sigma_\delta^2)$:} 
 \begin{equation}
 \mathcolor{black}{
   \mathbb{E}[\textbf{F}(\psi \pm \delta,\theta)] = \mathbb{E}[\textbf{F}(\psi,\theta)] \pm \mathbb{E}[\textbf{F}(\delta,0)] = \textbf{F}(\psi,\theta).}
\end{equation}
\subsection{Antenna Misalignment}

Antenna misalignment can be defined for two different scenarios. First, it occurs by the elevation disequilibrium in A2A link such that $d_{\textbf{u}_T} \hat{\textbf{u}} \neq d_{\textbf{u}_R} \hat{\textbf{u}}$ where $\hat{\textbf{u}}$ is the unit vector in the direction of the corresponding antenna. Second, independent from the altitude difference, misalignment occurs when an aerial node misaims the radiation field where UAV's actual orientation or attitude does not match the desired or commanded orientation such that $\tilde{\textbf{U}}_T^a \neq \tilde{\textbf{U}}_R^a$ for any $ \textrm{R}_z(\Theta)$ on $\textbf{U}_{T,R}$. Under the $\pm \delta$ misalignment for all planes, the SNR of this model is the following
\begin{align}
    \widetilde{\textrm{SNR}}^\phi_{\ell,k} &= \frac{P_T |h_{\ell,k}|^2 G_T(\theta \pm \delta_{T1}, \psi \pm \delta_{T2}) G_R (\theta \pm \delta_{R1}, \psi \pm \delta_{R2})}{\sigma^2}.
    \label{snr1}
\end{align}
\textcolor{black}{
The derivation of the next step of Eq. (\ref{snr1}) uses the property as following below:}
\begin{align}
       D_T(\theta_i, \psi_j) &= D_T(\theta_i, \psi_k) \forall i,j,k \\
       D_R(\theta_i, \psi_j) &= D_R(\theta_i, \psi_k) \forall i,j,k 
\end{align}
\textcolor{black}{
Therefore, $G(\theta_i, \psi_j) = G(\theta_i, \psi_k)$ if the radiation efficiency is $1$. Thus, $G(\theta_i,\psi_j + \delta_{T2}) = G(\theta_i,\psi_j)$ for both Tx and Rx. In this case, the neutral element of $\psi$ can be dropped from $\widetilde{\textrm{SNR}}^\phi_{\ell,k}$} which turns into the following:
\begin{align}
\widetilde{\textrm{SNR}}^\phi_{\ell,k} &= \frac{P_T |h_{\ell,k}|^2 G_T(\theta \pm \delta_{T1}) G_R (\theta \pm \delta_{R1})}{\sigma^2}.
\end{align}
That is, azimuth misalignment in BRF has no impact on the RSS level in an ideal short dipole while the elevation misalignment caused by both altitude and attitude mismatches have a direct relationship with the RSS and therefore the SNR. 

\subsection{Antenna with RF Impairments and Defects}
\label{secimpa}
Due to the fabrication \cite{ruze1952effect} and finite length antenna placement on UAV board \cite{badi2020experimentally, namiki2003improving},
antenna imperfections can be encountered and \textcolor{black}{these} flaws can be seen \textcolor{black}{in the} antenna field pattern. A comparative analysis was conducted between an ideal and non-ideal short dipole antenna, as illustrated in the Figure \ref{impairment}. Once the antenna response becomes \textcolor{black}{defective, the antenna possesses} some directional antenna characteristics such as main lobe, back lobe and side lobe levels (SLL). Furthermore, the HPBW intervals on elevation angle are not $[\frac{\pi}{4},\frac{-\pi}{4}]$ and $[\frac{3 \pi}{4},\frac{5 \pi}{4}]$ anymore. On top of this, $G_{\textrm{max}}$ depends on both planes in non-ideal antennas, unlike the ideal case where $G_{\textrm{max}}$ depends on the elevation cut solely.  
Under the previously defined Eq. (\ref{impairk}) defective antenna, an \textit{i.i.d.} multivariate AWGN of $\zeta$ for each plane has been defined as \textcolor{black}{follows} 
\begin{equation}
    D_{E,H} = 
    \begin{bmatrix}
           D_{E} \\
           D_{H} 
    \end{bmatrix},     \mu_{\zeta} = 
    \begin{bmatrix}
           D_H(\psi) \\
           D_E(\theta) 
    \end{bmatrix},     \Sigma_{\zeta} = 
    \begin{bmatrix}
           \sigma_\psi^2(\psi) & 0  \\
           0  & \sigma_\theta^2(\theta)
    \end{bmatrix}   
\end{equation}
In \textcolor{black}{the} following steps, $G(\psi, \theta)$ can be calculated without loss of generality. Note that $\zeta$ impairments in \textcolor{black}{the} antenna also aggregate the error in misalignment mentioned in \textcolor{black}{the} previous section. The probability distribution for $\zeta$ is simply,
\begin{equation}
    p(x;\mu_\zeta,\Sigma_\zeta)=\frac{1}{2\pi |\det(\Sigma_\zeta)|} \exp{\bigg(\frac{-1}{2}(x-\mu_\zeta)^\intercal \Sigma_\zeta^{-1} (x-\mu_\zeta)\bigg)}.
    \label{multivariate}
\end{equation}
As described in Section \ref{sec2}, $\zeta$ will be an additive random variable on $(0, 2\pi]$ $E$-plane and $(0, \pi]$ $H$-plane. 
With the modified HPBW for each plane, the sum directivity becomes  \cite{balanis2015antenna}
\begin{equation}
    \frac{1}{D_{\textrm{sum}}} = \frac{1}{2} \Bigr[\frac{1}{D_E} + \frac{1}{D_H}\Bigr], 
\end{equation}
where 
\begin{align}
    D_E \simeq \frac{1}{\frac{1}{2\log 2} \int_0^{\Omega_E/2}  \sin\theta d \theta} \simeq \frac{16 \log 2}{\Omega_E^2}, \label{29}\\
    D_H \simeq \frac{1}{\frac{1}{2\log 2} \int_0^{\Omega_H/2}  \sin\theta d \theta} \simeq \frac{16 \log 2}{\Omega_H^2}, \label{30}
\end{align}
where $\Omega_H$ and $\Omega_E$ are HPBW of corresponding planes in radian. Turning back to the example in Figure \ref{impairment}, under a Gaussian impairment for both planes as defined in Eq. (\ref{impairk}), a $20$ MHz centre frequency operating ideal antenna responses in Figure \ref{impairment}(a) can transform into the Figure \ref{impairment}(b). \textcolor{black}{In} this example, HPBW shrank from $90\degree$ to $76.8\degree$, \textcolor{black}{the} main lobe \textcolor{black}{acquired} $-5.55$ dB at $344\degree$ and \textcolor{black}{the} back lobe has $-7.68$ dB at $164\degree$. Note that \textcolor{black}{the} same field pattern flaws \textcolor{black}{are} valid for both transmitter and receiver antennas due to the antenna reciprocity. 
\section{Meta-learning Solutions}
\label{sec4}
The problem arises from the estimation of $\phi_R^{\psi,\theta}$ when $|\phi_T^{\theta}-\phi_R^{\theta}| \neq 0$ leads to $\hat{\phi}_R^{\theta} \neq \angle\max \{\hat{P}_R$\} due to the MPC and antenna imperfections under the assumption of $\hat{P}_R\xrightarrow{p} P_R$, yet the solution is possible for some $\tilde{\textbf{U}}^a_{(T,R)} = \textbf{R}^{ab} \textbf{U}^b_{(T,R)}$ exist. The non-zero correlation between $\tilde{\textbf{U}}^a_{(T,R)}$ and $P_R \in \textbf{P}_R$ under the defined perfectly estimated LoS channel aimed to be exploited by using few-shot learning. Due to the uniqueness of each A2A trajectory channel $\textbf{h}_{R_i}$, for each A2A link an $\mathcal{S}$ \textcolor{black}{exists} where $\textbf{U}_{{(T,R)}_i}^a \subset \mathcal{S}$ and power constraints of $\textbf{U}_{{(T,R)}}^a$, we aim to solve this problem with a limited dataset. Accordingly, the L2L approach is resorted to \textcolor{black}{employing} domain generalization technique to overcome this barrier. 
\subsection{Meta-Learning}
We utilize MAML with semantic segmentation network to estimate the $\phi_R^{\psi,\theta}$ by using estimated $P_R$ for a ${\textbf{A}}_R^a$ positioned UAV in space with $\tilde{\textbf{U}}_R^r$ direction. Thuswise, few-shot learning with $N$-way $K$-shot classification has been employed to utilize gradient descend. Here, a set of $K$ exemplary data samples are employed for fine-tuning the model during the evaluation process, aimed at achieving the task of classifying $N$ distinct classes.


In the search of finding the incident wave angles for more general Tx - Rx for UAV channels, the RSS distribution in $\mathcal{S}$ such as $p_{\mathcal{S}}(P_R)$ is not only $h_{R_i} \in \mathcal{S}, \forall{i}$ dependent, but also relies on $\textbf{R}^{ab}_i$ which has non-probabilistic distribution. Considering aforementioned antenna imperfections, analyzing such correlation between each $\phi_R^{\psi,\theta} \in \Phi_R^{\Psi,\Theta}$ and $p_{\mathcal{S}}(P_R)$ is vastly difficult that any task based learning algorithms will eventually fail, as neither using an experimental dataset nor utilizing deeper layers of neural networks cannot help. In \textcolor{black}{an} L2L approach, meta-learning is able to train over the $\mathcal{T}_i$ distribution to prioritize the learning convergence over the accuracy itself. Hence, instead of minimizing the loss function $\mathcal{L}(f_{{P}_R} (\mathcal{D}))$ to optimize the model parameter, MAML lowers the misclassification error for the sake of fast adaptation of the learning over sampled $\mathcal{T}_i$ distribution as following  
\begin{equation}
\label{meta}
        \hat{\mathcal{Q}}_{P_R} = \argmin_{\mathcal{Q}_{P_R}}  \mathbb{E}_{\mathcal{T}_i \sim p(\mathcal{T})} \mathcal{L}_{\mathcal{T}}(f_{(\mathcal{Q}_{P_R})}).
\end{equation}
\textcolor{black}{In the L2L perspective, the behavior of different data subsets can be learned rather than the behavior of samples belonging to a dataset. }The structure of the Eq. ($\ref{meta}$) will be explained broadly in Section \ref{optimization}. \textcolor{black}{
It is acknowledged that, under strong assumptions such as having an ample amount of data and a precise understanding of the radio environment's behavior, a standalone CNN model might be more robust than meta-models. However, this assumption is often unrealistic. Meta-learning, on the other hand, offers a pragmatic approach, delivering reliable and realistic results for challenging problems in a manner that aligns with realistic and explainable AI principles.
}

\subsubsection{An Upper Bound for ``$n$''}
\label{hoefsec}
In concern of convergence of this approach for the optimization of $f_{\mathcal{Q}_{P_R}} : \textbf{P}_R \rightarrow
\tilde{\Phi}_R^{\Psi,\Theta}$, an upper probability bound for MAML can be shown with Hoeffding's inequality. The meta classifier $f_{\mathcal{Q}_{P_R}} $ defined on a finite domain of $P_R$ variables and a random evaluation set of $\mathcal{D}_{\textbf{eval}}$ classified by the $f_{\mathcal{Q}_{P_R}}$. For each $P_R$, ${\phi_R^{\psi,\theta}}$ classification result with $\nu_{({\Phi_{R}^{\Psi,\Theta}}),i}, i \in \{1,...,n\}$. With the knowledge of \textcolor{black}{the} true error rate of the classifier ($\mathbb{E}(f_{\mathcal{Q}_{P_R}})$), \textcolor{black}{the success probability of classifier can be bounded as following:}
\begin{equation}
\Pr\Bigl(\sum_i \nu_{({\Phi_{R}^{\Psi,\Theta}}),i} - \mathbb{E}[\nu_{({\Phi_{R}^{\Psi,\Theta}})}] \geq \epsilon\Bigl) \leq e^{-\frac{2\epsilon^2}{\sum_{i=1}^n (b_i - a_i)^2}},
\end{equation}
where the $\epsilon \geq 0$ and $a_i \leq {P_R}_i \leq b_i$ almost surely. In meta classification, $\nu_{({\Phi_{R}^{\Psi,\Theta}}),i}$ has been defined such that $\nu_{({\Phi_{R}^{\Psi,\Theta}}),i} = 0$ if $f_{\mathcal{Q}_{P_R}}$ misclassify and $\nu_{({\Phi_{R}^{\Psi,\Theta}}),i} = 1$ in case classifier successes. Therefore, a Binomial random variable specified as $a_i = 0$ and $b_i = 1$ for all $i$.
Thereby, the Hoeffding Bound is the following
\begin{equation}
\Pr\Bigl(\sum_i \nu_{({\Phi_{R}^{\Psi,\Theta}}),i} - \mathbb{E}[\nu_{({\Phi_{R}^{\Psi,\Theta}})}] \geq \epsilon\Bigl) \leq \exp{\Bigl({\frac{-2 \epsilon^2}{n}}\Bigl)}. 
\label{hoef}
\end{equation}
Rearranging the Eq. (\ref{hoef}) results with \cite{abu2012learning}
\begin{equation}
\begin{split}
&\Pr\Bigl({\frac{1}{n}\sum_{i=1}^n\nu_{({\Phi_{R}^{\Psi,\Theta}}),i}} - \mathbb{E}[\nu_{({\Phi_{R}^{\Psi,\Theta}})}] \geq \epsilon\Bigl) \leq \exp{\Bigl({-2 \epsilon^2 n}\Bigl)},\\
=&\Pr\Bigl(\Bigl|{\frac{1}{n}\sum_{i=1}^n\nu_{({\Phi_{R}^{\Psi,\Theta}}),i}} - \mathbb{E}[\nu_{({\Phi_{R}^{\Psi,\Theta}})}] \geq \epsilon\Bigl|\Bigl) \leq 2 \exp{\Bigl({-2\epsilon^2 n }\Bigl)}. 
\end{split}
\label{hf}
\end{equation}
Following the Eq. (\ref{hf}), the minimum number of samples \textcolor{black}{required} for a given confidence interval ($\alpha_{\tilde{\Phi}_R}$) can be found. Therefore, the probability of accuracy error with $(\mathbb{E}(f_{\mathcal{Q}_{P_R}}) \pm \epsilon)$ confidence interval for the $\mathcal{L}(f_{{P}_R} (\mathcal{D}))$ optimized using $\{\mathcal{D}_{\textrm{train}}^{(i)},\mathcal{D}_{\textrm{test}}^{(i)}\}$ is the following
\begin{equation}
    \alpha_{\tilde{\Phi}_R} = \Pr(\nu_{({\Phi_{R}^{\Psi,\Theta}})}\in (\mathbb{E}(f_{\mathcal{Q}_{P_R}}) - \epsilon, \mathbb{E}(f_{\mathcal{Q}_{P_R}}) + \epsilon)) \leq 2 \exp{\Bigl({-2 \epsilon^2 n}\Bigl)}. 
\end{equation}
Finally, a tight bound for the $n$ is the following
\begin{equation}
    n \geq \frac{\log(2/\alpha_{\tilde{\Phi}_R})}{2 \epsilon^2}, \alpha_{\tilde{\Phi}_R} \neq 0.
\end{equation}
As a result, not only number of \textcolor{black}{samples} required can be estimated for a success threshold, but also the accountability of the model can be interpreted with the described bound.

\begin{figure*}[h] 
    \centering
    \includegraphics[width=0.95\textwidth]{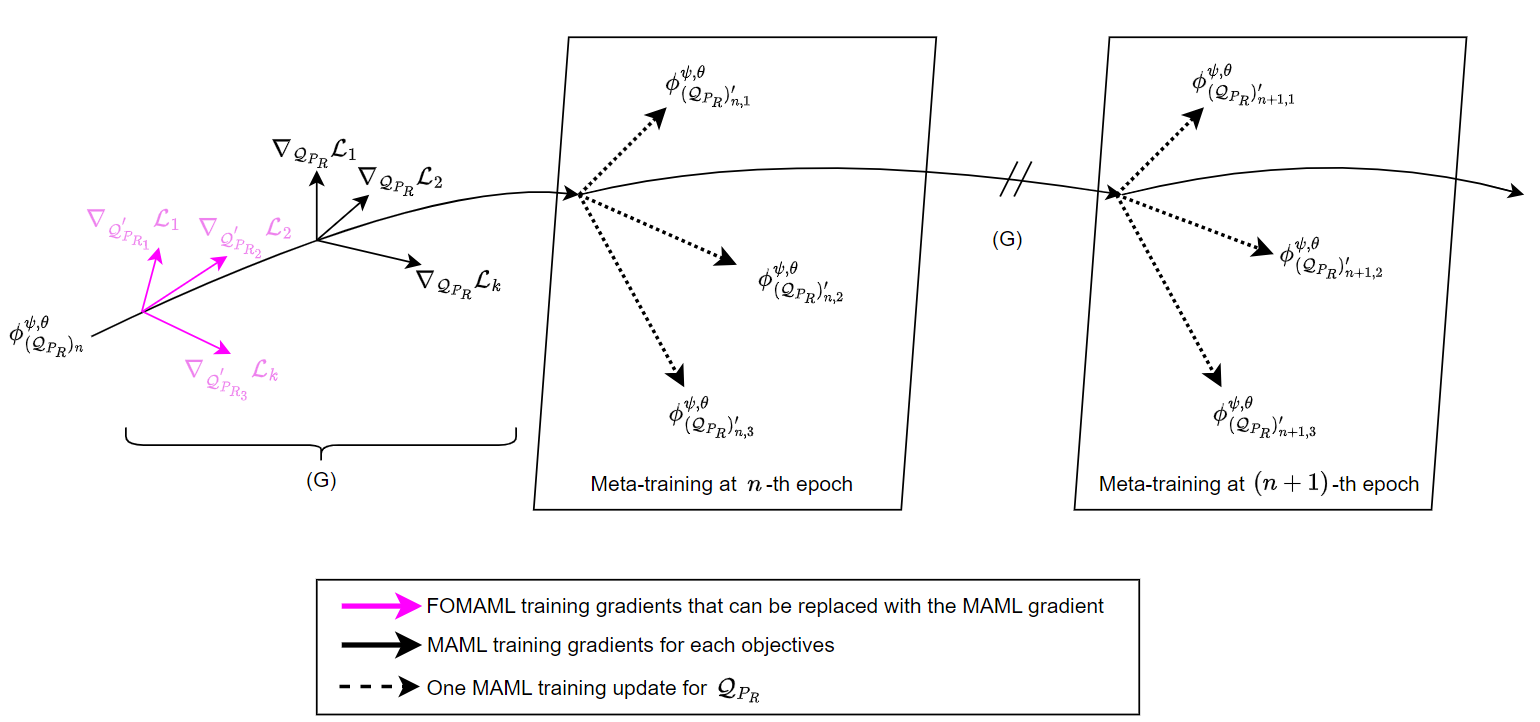}
    \caption{ $\phi^{\psi,\theta}_{(\mathcal{Q
}_{P_R})_n}$ adaptation for the optimal parameter vector for task $i$ in each iteration. $k$ objectives of sample beam angles for 3 inner steps has been shown for both MAML and FOMAML.}
    \label{flowchart}
\end{figure*}

\subsubsection{Task and Dataset Construction}
\label{dataset}
The task $\mathcal{T}_i$ associated with $\mathcal{D} = \{{\mathcal{D}_{\textrm{train}},\mathcal{D}_{\textrm{test}}}\}
$ where $\mathcal{D}_{\textrm{train}}$ is the support set used during the learning and $\mathcal{D}_{\textrm{test}}$ is the prediction set for training and testing. Each set contains feature vectors as $P_R^{M\times N} \in \textbf{P}_R$ as the set of corresponding incident wave angle $(\Phi_R^{\Psi,\Theta})^{T_1 \times T_2}$ as where $\textbf{P}_{R}^{(m)} \subset (\Phi_R^{\Psi(j),\Theta(k)}) \subset \kappa_{i}, m \in \{1,..,M\}, j\in\{1,..,T_1\}, k\in\{1,..,T_2\}, i\in\{1,..,I\}$. The following structure of $\mathcal{D}$ \textcolor{black}{allows} to demonstrate the full potential of MAML as following
\begin{equation}
    \begin{cases}
    \kappa_i \subset \mathcal{D}_{\textrm{train}}, & \text{if $i$ odd},\\
    \kappa_i \subset \mathcal{D}_{\textrm{test}} , & \text{otherwise},
    \end{cases}
\end{equation}
where $(i,j,k,m) \in \mathbb{Z}$. In this way, not only $\mathcal{D}_{\textrm{train}}$ distinct from the $\mathcal{D}_{\textrm{test}}$, but also $\mathcal{D}$ does not possess any other information such as SNR, CSI or $\textbf{F}_{T,R}$. Note that, the antenna with $\textbf{F}_{T,R}((\psi^{(i)}),\theta^{(j)}) = \textbf{F}_{T,R}((\psi^{(k \neq i)}),\theta^{(j)}), \forall i,j,k\in \left(0,\pi\right]$ allows constructing such a $\Phi_R^{\Psi^{(.)},\Theta^{(k)}}$ due to the fact that $p(P_R|\psi_R,\theta_R) = p({P}_R|\theta_R)$. Therefore, utilizing $\theta_R$ for the whole dataset drops the amount of used dataset from ($T_1 \times T_2$) to ($1 \times T_2$) without losing any information. Thus, $N$-way $K$-shot classification can be done with $K \subset \mathcal{D}_{\textrm{train}}$ samples from different $N$ classes. The defined dataset structure for \textcolor{black}{the} A2A link including antenna imperfections will be called 3D propagation model (\textit{3DPM}) in the remaining part of the study.

\textcolor{black}{Hereof, $\mathcal{D}$ is generated by utilizing the impairment modeling discussed in Section III and the derived channel modeling presented in Section II. Therefore, probability distribution of the far field constrained UAV distances with $\phi^{\theta}$ orientation follows $p_{\phi^\theta}(\phi^\theta|d)$ in Eq. (\ref{pdff}). The joint antenna impairments distribution on $E$ plane $H$ plane is given in Eq. (\ref{multivariate}), and the directivity with Gaussian impairments are given in Eq. (\ref{29}) and Eq. (\ref{30}).} Measured $P_R$ is the noisy measurement under $\textbf{h}_{\ell,k}$ with an ideal antenna that contains $\mathbb{U}\sim [0,1]$ PLF. Since the $\mathbb{E}(\textbf{F}_R (\psi,\theta,\zeta)) = \textbf{F}_R (\psi,\theta)$, non-idealities in antenna are not included in the $\mathcal{D}$. The same assumption holds for unbiased Gaussian misalignment where $\mathbb{E}(\delta_1) = \mathbb{E}(\delta_2) = 0$. \textcolor{black}{
 Recall that $k$ way transmission indicates the duplex direction between the UAVs, so that A2A nodes can follow two way communication model \cite{cabezas2023framework}. However, in terms of meta training data, it does not require any additional data since the duality of the states. This means that, every state of $\textbf{U}_{T}$ to $\textbf{U}_{R}$ of $k=1$ has an equivalence when $k=2$. For each }
$\mathcolor{black}{\textbf{p}_{T}^1 \xrightarrow[k=1]{h} \textbf{p}_{R}^2}$, \textcolor{black}{a link such} $\mathcolor{black}{\textbf{p}_{T}^2 \xrightarrow[k=2]{h} \textbf{p}_{R}^1}$ \textcolor{black}{exist}. Every supervised classification task can be generalized as following \cite{finn2017model}
\begin{equation}  
    \label{task}
    \mathcal{T}=\{\mathcal{L}(P_{R}^{(1)},\phi_R^{\theta,\psi,(1)}), p(P_{R}^{(1)}),p(P_{R}^{t+1}|P_{R}^{t},\phi_R^{\theta,\psi,(t)}),1\}.   
\end{equation}

In MAML, $K$-shot L2L model aimed to obtain a $\mathcal{T}_i \sim p(\mathcal{T}), \forall i \in K$, and then during the meta-training, $f_{\mathcal{Q}_{P_R}}$ trains with the $K$ samples by using the loss function $\mathcal{L}_{\mathcal{T}_i}$ that will be defined later on. As mentioned before, meta-test results with enhanced tasks of $\hat{\mathcal{T}}_i$, concluding with the more optimized $\hat{f}_{\mathcal{Q}_{P_R}}$ with $\hat{\mathcal{Q}}_{P_R}$ parameters. 
\begin{algorithm}
	\caption{MAML Model Optimization for Incident Wave Angle Estimation} 
    \label{alg}
	\begin{algorithmic}[1]
        \REQUIRE $p(\mathcal{T})$: distribution over RSS measurement tasks for different sets of $\textbf{h}_{\ell,k} $ (\textit{3DPM }dataset)
        \REQUIRE $\alpha$: Inner learning rate, $\beta$: Meta step size
        \STATE Randomly initialize $\mathcal{Q}_{P_R}$
        \WHILE {not done}
        \STATE Sample Meta-RSS batch measurements $\mathcal{T}_i$$ \sim$ $p(\mathcal{T})$ in Eq. (\ref{task})
        \FOR {\textbf{all} $\mathcal{T}_i$}
        \STATE Sample $K$ data points $\mathcal{D}$=$\{P_R^{(j)}, \phi^{\theta(j),\psi(\cdot)}\}$ from $\mathcal{T}_i$
        \STATE Evaluate $\nabla_{\mathcal{Q}_{P_R}} \mathcal{L}_{\mathcal{T}_i}(f_{\mathcal{Q}_{P_R}})$ using $\mathcal{D}$ and $\mathcal{L}_{\mathcal{T}_i}$ in Eq. (\ref{loss})
        \STATE Evaluation of tuned parameters with SGD: $(\mathcal{Q}_{P_R})'_i=\mathcal{Q}_{P_R}-\alpha \nabla_{\mathcal{Q}_{P_R}} \mathcal{L}_{\mathcal{T}_i}(f_{\mathcal{Q}_{P_R}})$ in Eq. (\ref{backpro})
        \STATE Sample data points $\mathcal{D}'_i$=$\{P_R^{(j)}, \phi^{\theta(j),\psi(\cdot)}\}$ from $\mathcal{T}_i$ for the meta-update.
        \ENDFOR
        \STATE Update $\mathcal{Q}_{P_R} \leftarrow \mathcal{Q}_{P_R}-\beta \nabla_{\mathcal{Q}_{P_R}} \sum_{\mathcal{T}_i \sim p(\mathcal{T})} \mathcal{L}_{\mathcal{T}_i}(f_{\mathcal{Q'}_{P_{R_{i}}})}$ using each $\mathcal{D}_i'$ and $\mathcal{L}_{\mathcal{T}_i}$ by Eq. (\ref{lossum}) and Eq. (\ref{18}) respectively.
        \ENDWHILE
	\end{algorithmic} 
\end{algorithm}
3DPM is a custom dataset for a supervised meta-learning algorithm generated with the discussed system model in Section II. It aims to convert a data learning process into task learning as efficiently as possible, so that the L2L approach can be performed. In the meta-learning, a training dataset $\mathcal{D}_{\textrm{train}}$ is used for the training process and a small test set $\mathcal{D}_{\textrm{test}}$ for adaptation to tune the parameters with each RSS received \cite{finn2017model}. Domain generalization can be summarized in two steps:
\begin{enumerate}
    \item Generate a set containing measured $P_R$ for $M$ different fading channel set by the $K-$factor.)
    \item Separate the generated set into two parts with respect to fading channel $K$- factor. So that, different instances of similar tasks (which is the wave angle estimation) can be meta-trained during $\mathcal{D}_{\textrm{test}}$ use.
    \item Use $n$ amount of task within $\mathcal{D}_{\textrm{train}}$ and $\mathcal{D}_{\textrm{test}}$ for the learning adaptation.  
\end{enumerate}
The first \textcolor{black}{step} ensures that the AI model is exposed to diverse datasets, which is preferable for any learning technique. The second and third \textcolor{black}{step} ensures that domain generalization is performed and $n$ number of tasks learned and $\mathcal{D}_{\textrm{test}}$ supports the $\mathcal{Q}_{P_R}$ for learning the already learned model.


\label{sec3}
\subsubsection{Meta-Optimization for Incident Wave Angle Estimation}
\label{optimization}
The mutual correlation for each $\textbf{P}_R \mathrel{\widehat{=}}
 \Phi_R^{\Psi,\Theta}$ optimization has been done by the tuning a model $\mathcal{Q}_{P_R}$ parameter in the negative direction of the gradient as following 
\begin{equation}
\label{sgd}
(\mathcal{Q}_{P_R})'_i= \mathcal{Q}_{P_R} -\alpha \nabla_{\mathcal{Q}_{P_R}} \mathcal{L}_{\mathcal{T}_i}^{(0)}(f_{\mathcal{Q}_{P_R}}),
\end{equation}
where $f_{\mathcal{Q}_{P_R}}$ is the model with parameters $\mathcal{Q}_{P_R}$, $\mathcal{T}_i$ is the task of optimization by utilizing the datasets $\{\mathcal{D}_{\textrm{train}}^{(i)},\mathcal{D}_{\textrm{test}}^{(i)}\}$ and $\mathcal{L}(\cdot)^{(i)}$ is the cross entropy loss function of $i-$th data batch for the supervised classification task $\mathcal{T}_i \sim p(\mathcal{T})$ defined as following
\begin{multline}
\label{loss}
\mathcal{L}_{\mathcal{T}_i}(f_{\mathcal{Q}_{P_R}}) = \sum_{\textbf{P}_R^{(j)},\Phi^{\Psi,\Theta(j)} \sim \mathcal{T}_i} \Phi^{\Psi,\Theta(j)} \log f_{\mathcal{Q}_{P_R}} (\textbf{P}_R^{(j)}) + \cdots \\
+ (1-\Phi^{\Psi,\Theta(j)}) \log (1-f_{\mathcal{Q}_{P_R}}(\textbf{P}_R^{(j)})), 
\end{multline}
where the loss function input/output pairs are denoted as $P_R^{M\times N} \in \textbf{P}_R$ and $\phi^{\psi,\theta} \in \Phi^{\Psi,\Theta}$ respectively. In such a scenario with initial parameter of $(\mathcal{Q}_{P_R})^{\circ}$ is selected, the inner gradient steps are following
\begin{equation}
\begin{split}
    &(\mathcal{Q}_{P_R})_0 = (\mathcal{Q}_{P_R})^{\circ}\\
    &(\mathcal{Q}_{P_R})_1 = (\mathcal{Q}_{P_R})_0 -\alpha \nabla_{\mathcal{Q}_{P_R}} \mathcal{L}_{\mathcal{T}_i}^{(0)}
    ((\mathcal{Q}_{P_R})_0)\\
    &(\mathcal{Q}_{P_R})_2 = ({\mathcal{Q}_{P_R}})_1 -\alpha \nabla_{\mathcal{Q}_{P_R}} \mathcal{L}_{\mathcal{T}_i}^{(0)}
    ((\mathcal{Q}_{P_R})_1)\\
    &\hspace{27.9mm}\vdots\\
    &(\mathcal{Q}_{P_R})_k = (\mathcal{Q}_{P_R})_{k-1} -\alpha \nabla_{\mathcal{Q}_{P_R}} \mathcal{L}_{\mathcal{T}_i}^{(0)}
    ((\mathcal{Q}_{P_R})_{k-1}).
    \label{backpro}
\end{split}
\end{equation}
The naturalization of meta-tasks to evaluate $\hat{\mathcal{Q}}_{P_R}$ by recursive iteration until to get an optimum $\dot{\phi}_R^{\psi,\theta}$ steps has been shown as 
\begin{equation}
    \begin{split}
    \hat{\mathcal{Q}}_{P_R} &= \argmin_{\mathcal{Q}_{P_R}}  \sum_{\mathcal{T}_i \sim p(\mathcal{T})} \mathcal{L}_{\mathcal{T}_i}^{(1)} \Bigl(f_{(\mathcal{Q}_{P_R})'_i}\Bigl), \\
     &=\argmin_{\mathcal{Q}_{P_R}} \sum_{\mathcal{T}_i \sim p(\mathcal{T})} \mathcal{L}_{\mathcal{T}_i}^{(1)}\Bigl(f_{\mathcal{Q}_{P_R}-\alpha \nabla_{\mathcal{Q}_{P_R}}\mathcal{L}_{\mathcal{T}_i}^{(0)}(f_{\mathcal{Q}_{P_R}})}\Bigl).
    \end{split}
    \label{lossum}
\end{equation}
Hence, the update of $\mathcal{Q}_{P_R}$ within the same task $\mathcal{T}$ is
\begin{equation}
\label{18}
    \mathcal{Q}_{P_R} \xleftarrow[\mathcal{T}_i]{} \mathcal{Q}_{P_R} - \beta \nabla_{\mathcal{Q}_{P_R}} \sum_{\mathcal{T}_i \sim p(\mathcal{T})} \mathcal{L}^{(1)} \Bigl(f_{\mathcal{Q}_{P_R} -\alpha \nabla_{\mathcal{Q}_{P_R}} \mathcal{L}^{(0)} (f_{\mathcal{Q}_{P_R}}) }\Bigl).
\end{equation}
As can be seen in Eq. (\ref{backpro}) and \textcolor{black}{Algorithm } \ref{alg}, SGD performs the second derivation on $\mathcal{Q}_{P_R}$ in the evaluation of $(\mathcal{Q}_{P_R})'_i$. As a significant note, the $(\mathcal{Q}_{P_R})'$ can be \textcolor{black}{achieved} with just a few steps in Eq. (\ref{backpro}), instead of calculating nearly 50 inner gradient steps as in \cite{yang2022meta} where the gradient vanishes and last layer does not update any weights. Using only the final gradient $k$ is an alternative approach for high dimensional calculations where the second-order derivatives \textcolor{black}{are} overcostly for trivial profit.

\textcolor{black}{
A meta-model can be further adapted to several fading channels. Here, we define the tasks in terms of RSS values under different fading channels. In this regard, the following two steps present the whole multi task meta-learning scheme:}
\begin{equation}
\mathcolor{black}{
    \hat{\mathcal{Q}}_{P_R} = \argmin_{\mathcal{Q}_{P_R}} \sum_{i=1}^n \mathcal{L}(\varphi_i, \mathcal{D}_{\textrm{test},i})} \end{equation}
\textcolor{black}{and}
\begin{equation}
    \mathcolor{black}{\varphi_i = f_{\mathcal{Q}_{P_R}}(\mathcal{D}_{\textrm{train},i})} 
\end{equation}

where $\mathcal{D}_{\textrm{train}}$ contains $\textbf{h}_{\ell, k}$, 
$\mathcal{D}_{\textrm{test}}$ contains ``few" amount of $\exists n: \textbf{h}_{\ell, k}$. In contrast to generic supervised learning, meta-learning minimizes the sum of $n$ task of the loss function on the $\mathcal{D}_{\textrm{test},i}$ with respect to $\varphi_i$. Thus, $n$ amount of tasks exist and each task $\tau_i$ owns $\mathcal{D}_{\textrm{train},i}$ and $\mathcal{D}_{\textrm{test},i}$. Here, $f_\mathcal{Q}$ is a neural network (which is a CNN in our case) and $\mathcal{D}_{\textrm{train},i}$ generates the $\varphi_i$ which can be used for $\mathcal{D}_{\textrm{test},i}$ within the same task that minimizes the loss. In this regard, the learning of RSS for $n$ amount of different fading channels is performed during the testing step for each task. Learning the generalized tasks in contrast to learning the dataset itself improves the adaptability of the model further. Thus, an idea over an unknown fading channel can be obtained.

\subsection{A Faster MAML Technique: FOMAML}

One alternative L2L variation to MAML \textcolor{black}{algorithm} is the FOMAML which \textcolor{black}{bypasses} the second derivative in the outer SGD. It has been shown that \cite{nichol2018first} FOMAML is not only more feasible to implement, but also it can generalize as \textcolor{black}{closely} as the original MAML in some situations where the dataset contains out-of-sample data points, as also known as outliers. In such cases, FOMAML with low order gradient calculation would \textcolor{black}{perform} even better.

Outer loop update of the $\mathcal{Q}_{P_R}$ in Algorithm \ref{alg}- step 10 for $k$ inner gradient step is simplified as taking the last inner gradient step ($k$) into consideration in Algorithm \ref{alg}- step 7. Thus, update equivalent disposes the $\nabla_{\mathcal{Q}_{P_R}}$ products within the inner step. Recalling that $\mathcal{L}$ is a differentiable function, Algorithm \ref{alg}- step 10 simply calculates the gradient of outer gradient $\mathcal{Q}_{P_R}$ as following 

\begin{equation}
\begin{split}
    \mathcal{Q}_{P_R} &\leftarrow \mathcal{Q}_{P_R}-\beta \nabla_{\mathcal{Q}_{P_R}} \sum_{\mathcal{T}_i \sim p(\mathcal{T})} \mathcal{L}_{\mathcal{T}_i}(f_{\mathcal{Q'}_{P_{R_{i}}}}) \\
    & \leftarrow \mathcal{Q}_{P_R}-\beta  \sum_{\mathcal{T}_i \sim p(\mathcal{T})} \nabla_{\mathcal{Q}_{P_R}} \mathcal{L}_{\mathcal{T}_i}(f_{\mathcal{Q'}_{P_{R_{i}}}}) \\
    & \leftarrow \mathcal{Q}_{P_R}-\beta  \sum_{\mathcal{T}_i \sim p(\mathcal{T})} (\nabla_{\mathcal{Q}_{P_R}} \mathcal{Q}'_{{P_{R_{i}}}}) \nabla_{\mathcal{Q}'_{{P_{R_i}}}} \mathcal{L}_{\mathcal{T}_i}(f_{\mathcal{Q}'_{P_{R_{i}}}})
    \end{split}
    \label{foma}
\end{equation}
Using the inner gradient equilevant of $(\mathcal{Q}_{P_R})_i'=\mathcal{Q}_{P_R}-\alpha \nabla_{\mathcal{Q}_{P_R}} \mathcal{L}_{\mathcal{T}_i}(f_{\mathcal{Q}_{P_R}})$ of Eq. (\ref{sgd}) on Eq. (\ref{foma})  finalizes the objective update analysis for MAML as in \small

\begin{equation}
  \mathcal{Q}_{P_R} \leftarrow \mathcal{Q}_{P_R}-\beta \sum_{{\mathcal{T}_i} \sim p(\mathcal{T})} \underbrace{(\textbf{I}-\alpha  \nabla^2_{\mathcal{Q}_{P_R}} \mathcal{L}_{\mathcal{T}_i}(f_{\mathcal{Q}_{P_R}})}_{\textrm{ Requires Hessian}}  \nabla_{\mathcal{Q}'_{P_{R_i}}} \mathcal{L}_{\mathcal{T}_i}(f_{\mathcal{Q'}_{P_{R_{i}}}}).
  \label{openouter}
\end{equation}
\normalsize Thus, the explicit expression of optimizer gradient given in Eq. (\ref{18}) is obtained in Eq. (\ref{openouter}). FOMAML algorithm simply ignores the Hessian term and takes the indicated term into account as $\textbf{I}$ for the current sample batch. This \textcolor{black}{technique} can be replaced with MAML as illustrated in Figure \ref{flowchart}.

\subsection{Incident Wave Azimuth Angle Estimation}

The solution for $\phi_R^{\psi}$ can be obtained by exploiting the invariant null zones of the antenna and the estimated elevation angle of \textcolor{black}{the} incident wave. Beam scanning, steering and eigendecomposition of \textcolor{black}{the} received signal via multiple signal \textcolor{black}{identification} classification (MUSIC) are some of the widely adopted techniques for achieving $\phi_R^{\psi}$ estimation objective. Without needing any CSI or MPC information, $\phi_R^{\psi}$ can be found directly by the $P_R$ and estimated $\tilde{\phi}_R^{\theta}$ by few-shot learning. Therefore, for this part of the study, $\phi_R^{\psi}$ can be found by rotating the antenna pattern for each $\phi_R^{\theta}$ until to find the minimum RSS level, as will be indicated as the $\phi_R^{\psi}$. With the assumption that $\tilde{\phi}_R^{\theta}$ is correct, a similar approach in \cite{malajner2011angle}, $\tilde{\phi}_R^{\psi}$ solution is possible in $\tilde{\phi}_R^{\theta}$ direction. Holding the previous assumption that Rx antenna has the same $\textbf{A}^r_R$ with its UAV, this rotation needs to be done with \textcolor{black}{the} kinematic motion of the $\textbf{U}_R^r$. First, locate the antenna along \textcolor{black}{the} axis through the elevation angle by either\footnote{The difference in roll and pitch initials corresponds to different $\psi$ angles, therefore an arbitrary \textcolor{black}{choice} does not affect the $\psi$ estimation since steering covers each of beam samples.} roll or pitch orientation as \textcolor{black}{follows}
\begin{equation}
\textbf{U}_\textrm{steer}^\theta = \textrm{R}_z(\textbf{I}) \textrm{R}_y(\Theta_2(\tilde{\psi}^\theta)) \textrm{R}_x(\textbf{I}) \textbf{U}_b,
\end{equation}
where the $\tilde{\psi}^\theta$ is the rotation angle through the estimated $\theta$ by the MAML. Beam steering with $l \in \textbf{l}$ steps are defined for each beam samples as following

\begin{equation}
    P_R^{\theta,\psi(l)}=
    \begin{cases}
        \textrm{R}_z(\Theta_1(l)) \textrm{R}_y(\textbf{I}) \textrm{R}_x(\textbf{I}) \textbf{U}_\textrm{steer}^\theta, & \text{if $\hat{\textbf{u}}_0$ $\|$ $\hat{e}_z$} \\
        \textrm{R}_z(\textbf{I}) \textrm{R}_y(\Theta_2(l)) \textrm{R}_x(\textbf{I}) \textbf{U}_\textrm{steer}^\theta, & \text{if $\hat{\textbf{u}}_0$ $\|$ $\hat{e}_y$} \\
        \textrm{R}_z(\textbf{I}) \textrm{R}_y(\textbf{I}) \textrm{R}_x(\Theta_3(l)) \textbf{U}_\textrm{steer}^\theta, & \text{if $\hat{\textbf{u}}_0$ $\|$ $\hat{e}_x$}
    \end{cases}
\end{equation}
Sweeping the beams so that the optimum $\phi^\psi$ for the corresponding $\phi^\theta$ will be the $l$-th beam sample that gets the minimum $P_R$, \textcolor{black}{ideally} zero.
\begin{equation}
    \tilde{\phi}^\psi = \argmin_{l\in \textbf{l}} P_R^{\theta,\psi(l)} 
\end{equation}
This concludes the both azimuth and elevation angle estimations of an incident wave. 


\begin{algorithm}
	\caption{CNN Model Optimization for Incident Wave Angle Estimation} 
    \label{alg2}
	\begin{algorithmic}[1]
        \REQUIRE $\mathcal{T}$: RSS measurement tasks
        \REQUIRE Optimizer learning rate $(\alpha)$, Batch size ($b$)
        \STATE Randomly initiliaze $\mathcal{Q}_{P_R}$
        \STATE Randomly initiliaze task selection with $\mathcal{T}_i$$ \sim$ $\mathcal{T}$
        \FOR {\textbf{all} $\mathcal{T}_i$}
        \STATE Construction of training batch $\mathcal{B}^b_i$ with $\mathcal{D}$=$\{P_R^{(j)}, \phi^{\theta(j),\psi(.)}\}$ from $\mathcal{T}_i$
        \STATE Evaluate $\nabla_{\mathcal{Q}_{P_R}} \mathcal{L}_{\mathcal{T}_i}(f_{\mathcal{Q}_{P_R}})$ using $\mathcal{B}^b_i$
        \STATE Evaluation of tuned parameters with SGD: $\mathcal{Q}_{P_{R_i}}'=\mathcal{Q}_{P_R}-\alpha \nabla_{\mathcal{Q}_{P_R}} \mathcal{L}_{\mathcal{T}_i}(f_{\mathcal{Q}_{P_R}})$
        \ENDFOR
	\end{algorithmic} 
\end{algorithm}

\section{Numerical Results}
\label{secnum}
In this section, the numerical evaluations of the proposed methods have been presented and discussed. \textcolor{black}{The corresponding model parameters are placed \textcolor{black}{in} Table I.} Firstly, the simulation environments and performance metrics have been explained. Then, the impact of a non-ideal antenna radiation pattern on the fading channel has been \textcolor{black}{numerically} evaluated. In continuation, explained MAML and FOMAML algorithms have been implemented for the estimation of $\phi_R^{\psi,\theta}$. Lastly, a well-known CNN model has been generated for benchmark \textcolor{black}{tests} and comparison \textcolor{black}{tools}. \textcolor{black}{Due to use of model-agnostic approach, any model can be utilized along with the meta-optimization. For instance, it can be a support vector machine, decision tree, or a neural network (which is a simple CNN in our approach). That's why we utilize the exact same CNN model for a comparison, highlighting the significance of MAML improvement on the simple task optimization.}

\begin{figure}[] 
    \centering
    \includegraphics[width=0.5\textwidth]{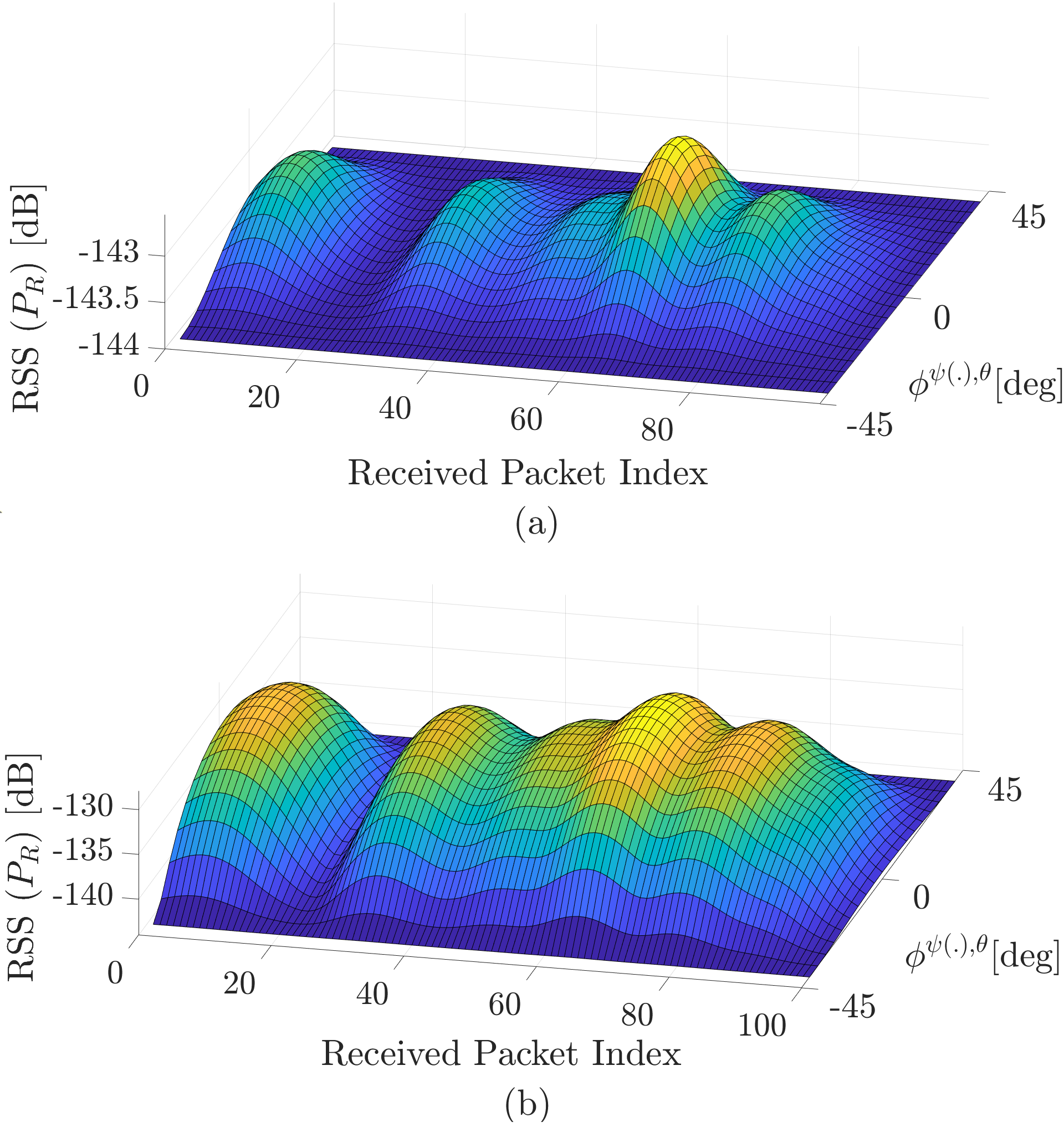}
    \caption{Impact of incident elevation angle of the signal on RSS level with $P_T=10$ dB (a) and $P_T=30 $ dB (b). }
    \label{gain}
\end{figure}

\textcolor{black}{Similary, a feed-forward $\mathrm{DenseNet}$ is utilized for the same objective with using the same dataset. The $\mathrm{DenseNet-BC}$ architecture with $k=12$ growth rate and $L=100$ layers scheme is summarized in Table \ref{dense}. It uses bottleneck layers in each Dense Block for parameter reduction. While the same optimizer momentum rate and loss function structure are utilized for supervised classification, specific emphasis is placed on highlighting the distinctive parameters unique to the $\mathrm{DenseNet}$ architecture. The model contains $3$ dense block which contains $96$ bottleneck layers where each layer takes all preceding feature maps as input. The drop rate is $0.2$, and total of contains $2$ transition layers between the dense blocks. The model progressively reduces the spatial dimensions of feature maps while increasing the number of channels, resulting in a compact representation suitable for beam classification. Therefore, the feature maps are compressed with the order of compression factor $C^* = 0.5$ within each transition layers. After the dense blocks, a global average pooling layer followed by softmax activation as a logistic classifier, just as in MAML.}

$M=20$ and $N=100$ sized RSS matrix as the set of $\mathcal{D}$ has been used to perform joint training and testing process. In previously defined 3DPM dataset, it has been set to $T_2=20$ instances of $180$ RSS samples that corresponds to each $\Phi^{\Theta(i)}_R$ for each $i\in[-\frac{\pi}{2}, \frac{\pi}{2}]$ from $\kappa = \{0,1,2,..,30\}$ channels. \textcolor{black}{One reason for using Rician fading with both high and low $\kappa$ is to create a $\mathcal{D}$ such that a base meta-model that utilizes $\mathcal{D}_{\text{train}} \subset \mathcal{D}$ can also excel in untrained and unseen tasks with fine-tuning parameters with $\mathcal{D}_{\text{test}} \subset \mathcal{D}$. Therefore, we turn the objective from Rician fading channel learning into \underline{a} fading channel learning.} \textcolor{black}{Diversity of prior distributions across the tasks is provided with channel fading \cite{park2020meta} as in Eq. (\ref{backpro}) for $\mathcal{D}_{\textrm{train}}$ and $\mathcal{D}_{\textrm{test}}$ subsets \cite{kumar2023effect}}.
There is no further augmentation and modification on the dataset. Throughout the numerical evaluation \textcolor{black}{of} the receiver, it has been assumed that CSI at the receiver is known. The different $K$ and $N$ scenarios of meta-classification have been tried within the same dataset. The impact of $N$-way classification has been evaluated with $N=\{6,8,10\}$ while the $K$-shot effect has been compared within the same $N$-way classification. In these concepts, the evaluation of meta-classification steps follows the Algorithm \ref{alg} step $4-9$ as \textcolor{black}{follows}: 
\begin{enumerate}
    \item Select random $N$ classes from $\mathcal{D}_{\textbf{eval}}$.
    \item Pick $K$ instances for each class to fit the model.
    \item Performing the evaluation for different instances within the same class. 
\end{enumerate}
\textcolor{black}{Hence, the evaluation of meta-model performs SGD over $K$ samples to tune $\mathcal{Q}_{P_R}$ parameters, as the underlying principle of L2L that learning over the learned model.} The evaluation dataset $\mathcal{D}_{\textrm{eval}}$ contains $P_R^{1000 \times 100} \in \textbf{P}_{R,\textrm{eval}}$ generated on the 3DPM defined in Table \ref{3DPro} by the same method has been used to generate $\mathcal{D}$ with $M=1000$ in Section (\ref{dataset}).
The architecture model \textcolor{black}{consists of} $4$ of convolution \textcolor{black}{filter} blocks with each containing $64$ filters with $3\times 3$ kernel, a batch normalization, followed by a ReLU function and $2 \times 2$ max-pooling respectively. Each $P_R^{1\times 100}$ \textcolor{black}{sample} create $\textbf{P}_R$ and last layer activation function SoftMax returns an array of probability score as usual. Some key parameters of the MAML \textcolor{black}{have} been shown in Table \ref{3DPro}. The FOMAML follows the exact same process with the induced outer gradient version as explained previously. 
\subsection{Performance Metrics and Simulation Environment}

Accuracy is one of the metrics in evaluation of meta-classification defined as following
\begin{equation}
    \textrm{Accuracy} = \frac{\sum\limits^{\mathcal{D}_{\textrm{test}}} \mathbbm{1}\{\tilde{\textbf{c}}^{\phi,\theta}_R = \textbf{c}^{\phi,\theta}_R \}} {M\times N}
\end{equation}
where the $\tilde{\textbf{c}}^{\phi,\theta}_R$ is the estimated beam sample and $\textbf{c}^{\phi,\theta}_R$ is the ground truth beam sample. The very same accuracy definition is also valid for the evaluation with its dataset will be discussed. 

\textcolor{black}{Apart from the beam classification accuracy, the direct angle estimation errors can be bounded as following: 
Angle estimation error ($\epsilon_\phi = \phi - \overline{\phi}$) can be defined in two different categories: 
\begin{enumerate}
    \item Angle estimation error in case of correct beam prediction, denoted as $\epsilon_\phi^1$
    \item Angle estimation error in case of faulty beam prediction, denoted as $\epsilon_\phi^0$
\end{enumerate}
We note that $\epsilon_\phi^1 \leq \frac{\pi}{N} \leq \epsilon_\phi^0$ and 
$\epsilon_\phi^1 \sim \mathbb{U}(0,\frac{\pi}{N})$ and $\epsilon_\phi^0 \in (\frac{\pi}{N},\pi]$. Note that the resolution on angle estimation error can go down to $0$, as $N$ goes to infinity. In this case, if the classification is correct, the $\phi^{\psi, \theta}$ is correct with $\epsilon_\phi^1 = 0$. However, as we also discuss in the original manuscript, this is only possible with a classification accuracy trade-off.}

As indicated earlier, the classification error can be skeptical in Meta-RSS, therefore Hoeffding's PoC bounds the success point probability of the generated models as derived in Section \ref{hoefsec}.

\begin{align}
    \textrm{PoC} &=\exp{\Bigl({-2 \epsilon^2 n}\Bigl)}\\
    \textrm{Point Probability} &= \Pr\Bigl({\frac{1}{n}\sum_{i=1}^n\nu_{({\Phi_{R}^{\Psi,\Theta}}),i}} - \mathbb{E}[\nu_{({\Phi_{R}^{\Psi,\Theta}})}] \geq \epsilon\Bigl)
\end{align}

\subsection{Evaluation of the Results}
\begin{figure}[] 
    \includegraphics[width=0.5\textwidth]{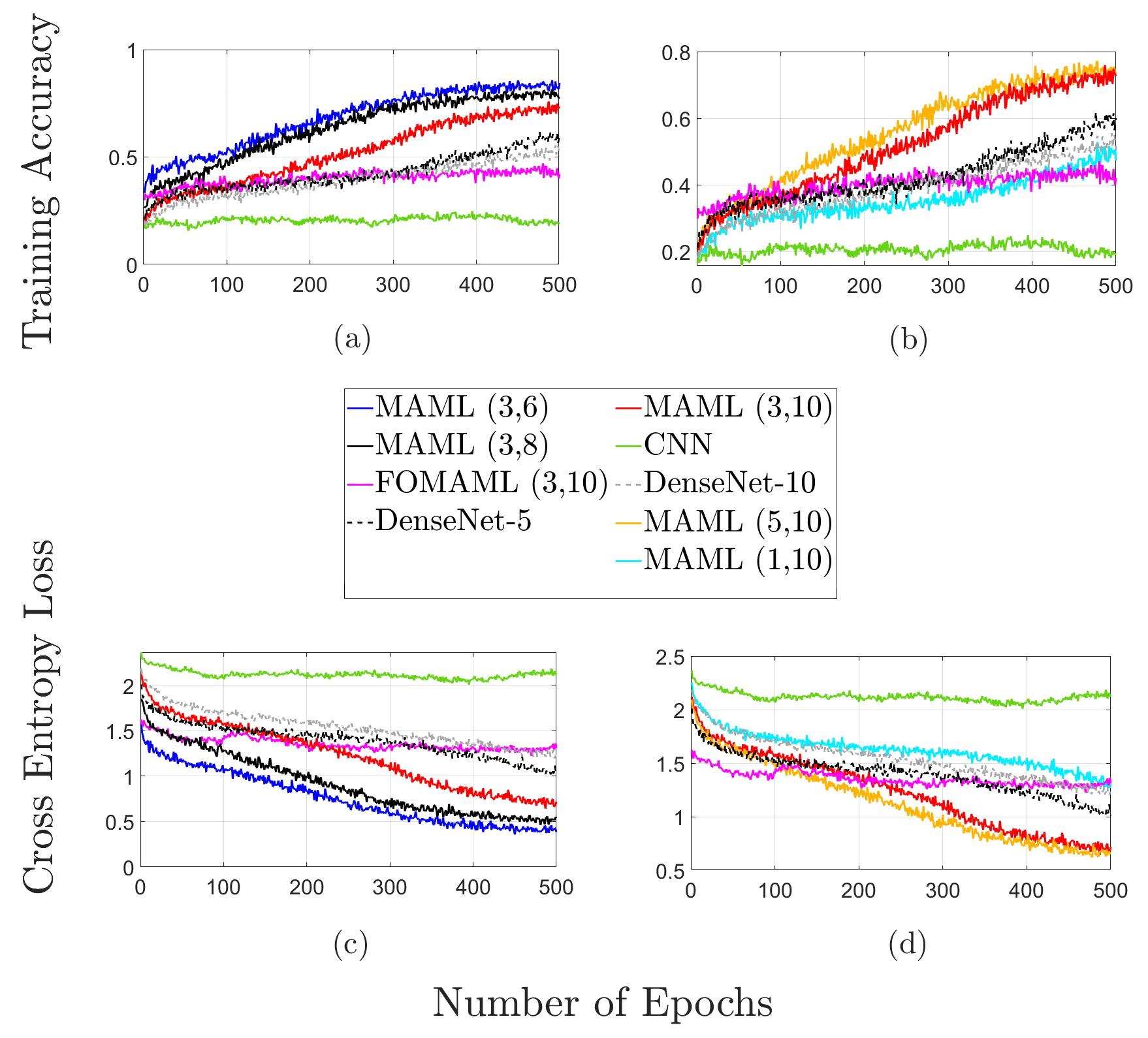}
    \caption{\color{black}{Learning curves comparison for different MAML($K$,$N$), FOMAML($K$,$N$), CNN and DenseNet-$N$ estimator models with the same iteration number.}}
    \label{learn}
\end{figure}
The impact of the antenna pattern has been illustrated in Figure \ref{gain}. The highlight of this study emphasized with the $P_R^{1\times 100}$ measurements over $\tilde{\textbf{U}}^r_R$ having an ideal antenna on $xy$- plane with $\phi^{\psi(\cdot),\theta} \in [-\pi/4,+\pi/4]$ incident elevation angle at the receiver. The increment in $P_T$ for $\textbf{U}^r_T$ demonstrates a widespread influence on RSS. However, the presence of a multiplicative incident wave angle term within the path loss restricts this influence regardless of the prevailing channel state. \textcolor{black}{In} the same figure, it can be seen that increment in $P_T$ becomes negligible for near-null pattern zones, showing the significance of the 3DPM involving antenna characteristics where the incident wave angle information can mitigate this $P_T$ bottleneck for the link. 

The loss and training accuracy curves of the generated models can be seen \textcolor{black}{in} Figure \ref{learn}, \textcolor{black}{which underscores the importance of MAML's enhancements in optimizing simple tasks, demonstrating successful generalization across $\mathcal{T}_i$ and transforming the basic CNN model into a robust and adaptive model. However, it is observed that, both $\mathrm{DenseNet-5}$ and $\mathrm{DenseNet-10}$ accomplished to capture features on 3DPM within the limited iteration allowance. 
} The NVIDIA Tesla-T4 has been utilized to perform learning algorithms with CUDA 12.0 in 6 hours. PyTorch framework \textcolor{black}{has} been used to perform few-shot learning algorithm to joint training-testing process. In FOMAML, the duration of training drops from $6$ to $2$ hours and a half hour with fast training convergence. The comparison of different $N$ with \textcolor{black}{the} same $K=3$ can be seen in Figure \ref{learn}(a) and (c), while the comparison of different $K$ with the same $N=10$ is shown in Figure \ref{learn}(b) and (d). With the $500$ epochs iteration, the models have reached their achievable performances where \textcolor{black}{they get bottlenecked} by the number of channels and dataset size, except MAML($5$,$10$). It depicts that increasing $N$ requires more iteration to converge as can be seen on MAML($5$,$10$) case. \textcolor{black}{Additionally, the simulations exclusively focus on examining the $\textbf{U}^r_R$ under rotational impact for the sake of simplicity. This deliberate choice positions the $\textbf{U}^r_R$ as the controlled variable, while the $\textbf{U}^r_T$ serves as a background to compare the impact of the receiver UAV state.}


\setlength\extrarowheight{4pt}

\begin{table}[t]
\centering
\caption{DenseNet Architecture}
\label{dense}
\resizebox{0.75\columnwidth}{!}{%
\begin{tabular}{lll}
\textbf{Layers} & \textbf{DenseNet}                                                  & \textbf{Output Size} \\ \hline \hline     
Convolution        & $3 \times 3$ & $32 \times 32$ \\ \hline
Dense Block 1            & $\Big[\begin{smallmatrix}1 \times 1 \\3 \times 3\end{smallmatrix}\Big] \times 24$   & $32 \times 32$ \\ \hline
\multirow{2}{*}{Transition Layer 1} & $1 \times 1$ conv & $32 \times 32$ \\ 
&$2\times 2$ avg pool & $16 \times 16$ \\ \hline
Dense Block 2         & $\Bigl[\begin{smallmatrix}1 \times 1 \\3 \times 3\end{smallmatrix}\Bigl] \times 108$   & $16 \times 16$ \\ \hline
\multirow{2}{*}{Transition Layer 2} & $1 \times 1$ conv & $16 \times 16$ \\ 
&$2\times 2$ avg pool & $8 \times 8$ \\ \hline
Dense Block 3           & $\Big[\begin{smallmatrix}1 \times 1 \\3 \times 3\end{smallmatrix}\Big]\times 150 $   & $8 \times 8$ \\ \hline
\multirow{2}{*}{Classification} & global avg. pool & $1 \times 1$ \\ 
&softmax & $10$     \\ \hline \hline                    
\end{tabular}%
}
\end{table}
\setlength\extrarowheight{0pt}
\begin{table}[b]
\caption{Few-shot learning and 3DPM parameters}
\resizebox{\columnwidth}{!}{
\begin{tabular}{lr||lr}

\textbf{Meta-Parameters }                                                   & \textbf{Value}      & \begin{tabular}[c]{@{}l@{}}\textbf{3D Propagation}\\\textbf{Model Parameters}\end{tabular} & \textbf{Value}                           \\ \hline \hline
\begin{tabular}[c]{@{}l@{}}Number of classes\\  ($N$)\end{tabular} & \{6,8,10\} & $\kappa$                                                                   & 12                              \\
Number of shots ($K$)                                              & \{1,3,5\}  & PLF                                                                        & $\mathbb{U}\sim [0,0.5]$                         \\
Inner train step ($k$)                                             & 1,3          & $\mu_\zeta$                                                                & 0                               \\
Inner learning rate ($\alpha$)                                     & 0.4        & $\Sigma_\zeta$                                                             & 0.2 \textbf{I} \\
Meta-RSS Batch size                                                & 10         & $\delta_{T1}, \delta_{R1}$ {[}deg{]}                                             & 0.05, 0.05                      \\
Number of Epochs                                                   & 500        & SNR {[}dB{]}                                                               & 0-30 dB                         \\
Task (outer) learning rate                                         & 0.001      & $d_x, d_y, d_z$ {[}m{]}                                                    & $\{100,10,5\}$                  \\
Meta step size ($\beta$)                                           & 0.999      & $\textbf{U}^a_T$,  $\textbf{U}^a_R$                                                          & $[0,0,1]^\intercal$                     \\ \hline \hline
\end{tabular}
}
\label{3DPro}
\end{table}

\begin{figure}[t] 
    \centering
    \includegraphics[width=0.5\textwidth]{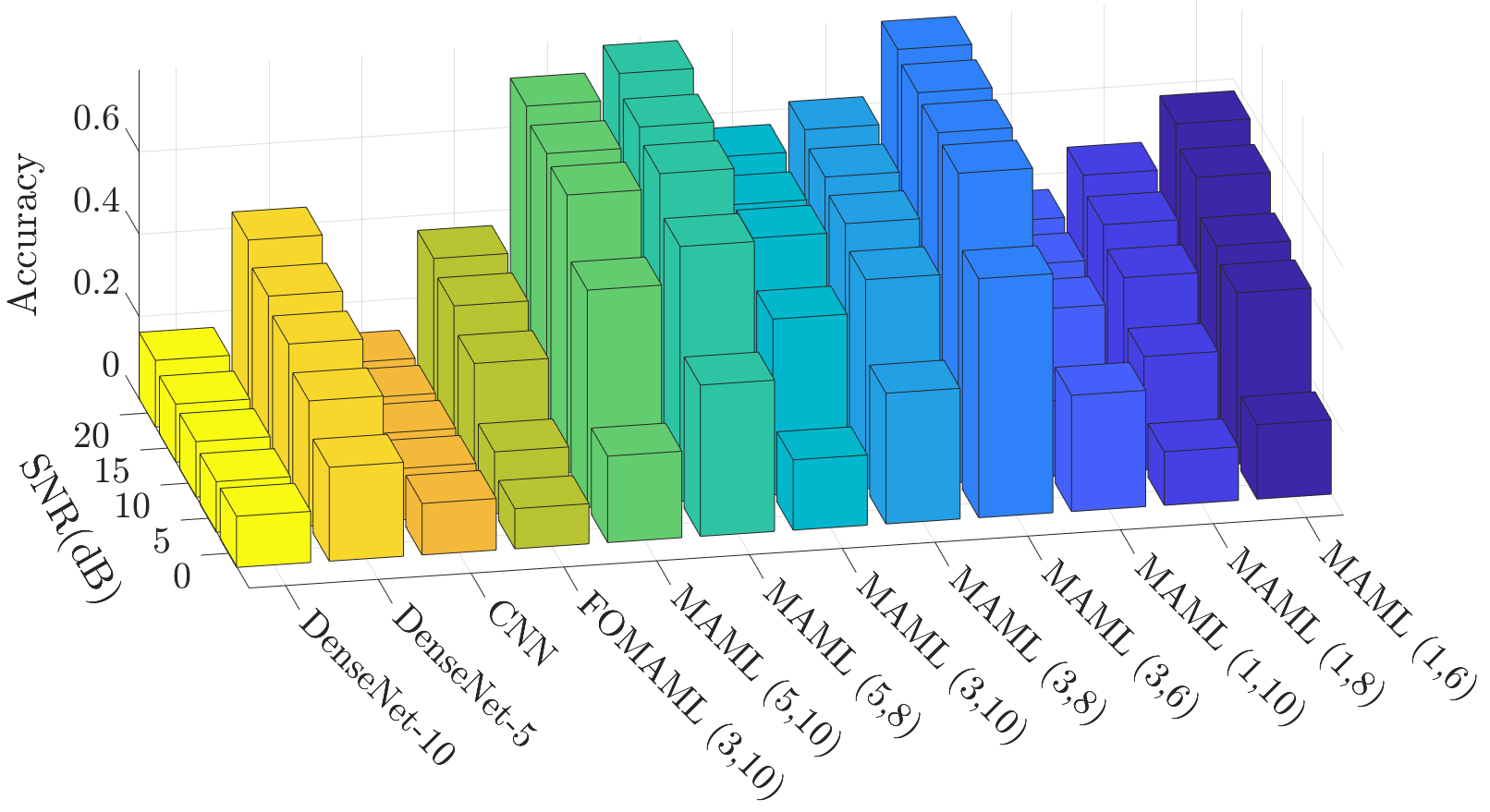}
    \caption{\color{black}{Evaluation of elevation classification with CNN, DenseNet-$N$, MAML($K$,$N$) and FOMAML($K$,$N$) models for $0-20$ dB SNR range. }}
    \label{eva}
\end{figure}

The $\phi^{\theta}_R$ estimation accuracy performance of different methods have been shown in Figure \ref{eva}. The initial observation reveals that the performance of the CNN classifier does not portray improvement as the SNR increases. The CNN, which estimates the beam samples with a high variance of approximately $0.16$, demonstrates overfitting and exhibits weakness compared to all other configurations. 

\textcolor{black}{Despite its learning capabilities, the evaluation of $\mathrm{DenseNet}$ does not accurately reflect its learning accuracy, as anticipated. Note that this outcome arises not only because $\mathcal{D}_{\textrm{eval}}$ includes unseen $P_R$ for $\phi^{\theta,\psi}$, but also because the objective itself is unseen, given that the evaluation fading channel is not part of $\mathcal{D}_{\textrm{train}}$. However, unlike CNN, both $\mathrm{DenseNet}$ architectures exhibit enhancements as the SNR level increases. For instance, $\mathrm{DenseNet-5}$ demonstrates a notable improvement, with its accuracy rising from $22.8$\% at $0$ dB to $43.9$\% at $20$ dB. However, this enhancement is less pronounced in $\mathrm{DenseNet-10}$.} Conversely, all other models consistently enhance their accuracy above the $0$ dB threshold. Notably, reducing the $N$ or increasing the $K$ consistently enhances the performance. After reaching a $15$ dB level, all scenarios encounter an error floor in different levels, with the MAML(3,6) configuration achieving a maximum accuracy performance of $0.74$. The MAML(1,10) experiment initially resembles the behavior of the CNN at $0$ dB but reaches to maximum accuracy performance of $0.3$ eventually. In general, the one-shot experiments demonstrate relative success but rather poor performance compared to the cases with $K=3$ and $K=5$. Table \ref{tab4} presents the classifier accuracy within the same SNR level for \textcolor{black}{different} model configurations.

\begin{figure}[t] 
    \centering
    \includegraphics[width=0.5\textwidth]{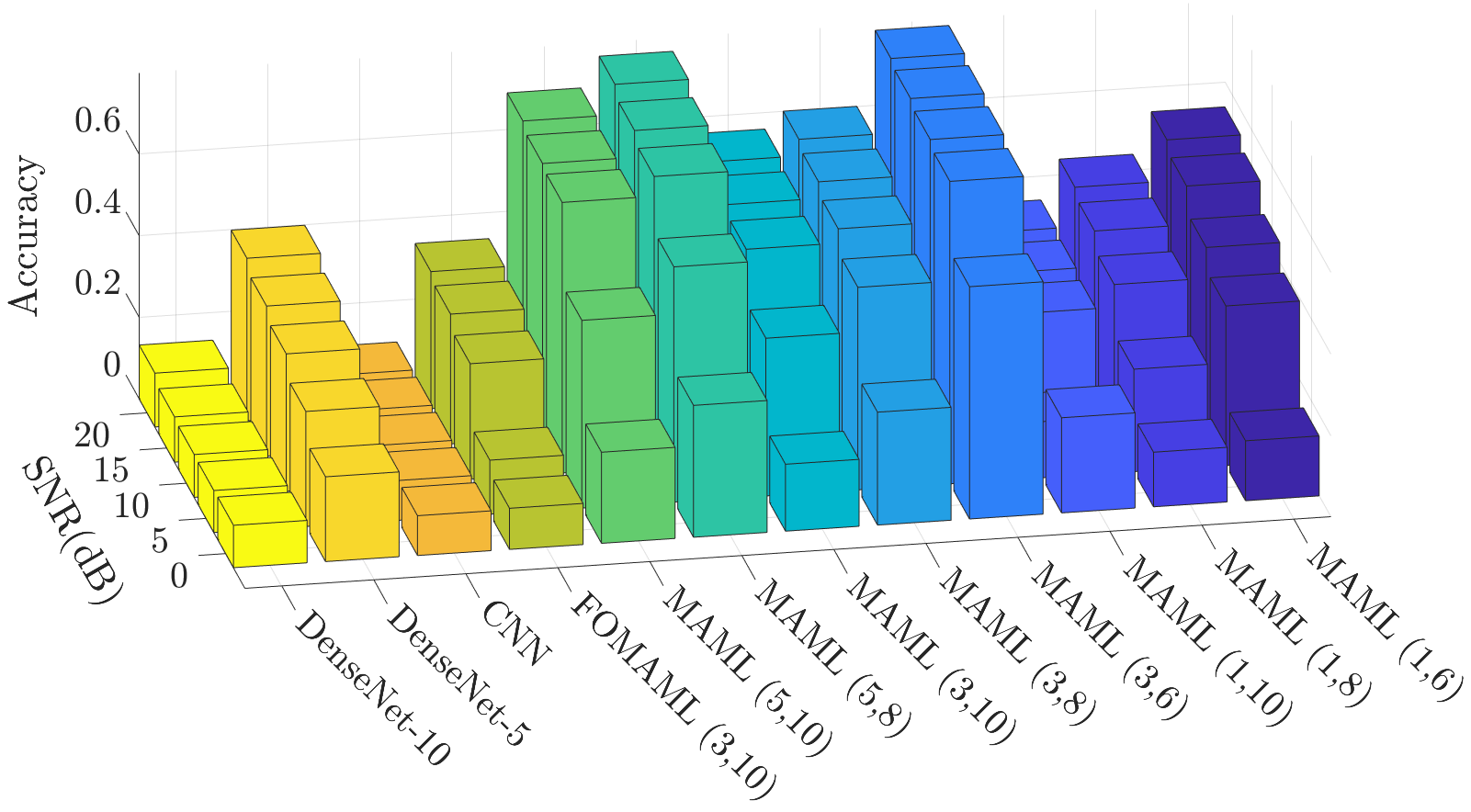}
    \caption{\color{black}{Evaluation of azimuth classification with CNN, DenseNet-$N$, MAML($K$,$N$) and FOMAML($K$,$N$) models for $0-20$ dB SNR range.}}
    \label{evb}
\end{figure}

Later on, the $\phi^{\psi}_R$ estimation accuracy for different cases \textcolor{black}{has} been evaluated in Figure \ref{evb}. Similarly, MAML and FOMAML \textcolor{black}{outperform} both state-of-the-art CNN and $\mathrm{DenseNet}$ models, not only in terms of accuracy but also in the convergence of the algorithms. It has been observed that the first step elevation estimation correlates with the azimuth estimation. Numerical observations indicate that while CNN \textcolor{black}{exhibits} a learning curve, it demonstrates suboptimal classification performance in evaluation tests. \textcolor{black}{Moreover, $\mathrm{DenseNet}-5$ shows similar trend with elevation accuracy. However, $\mathrm{DenseNet}-10$ performs as well as CNN and it loses the demonstrated improvement depending on the SNR.}

\textcolor{black}{
For comparison with previous researches of \cite{al2017two}, \cite{wu2019multiemitter} with similar purposes, our proposed method utilizes the omnidirectivity within $\phi^\theta \in (0, 2\pi]$ and $\phi^\psi \in (0, \pi]$, therefore only estimation loss appear through the nulls where $\theta = \{-\pi/2,\pi/2\}$. Understandably, the accuracy through the null regions lowers down due to lowered SNR. At this point, the advantage of A2A link can appear where the UAVs are free to restrict their orientation and position to meet the controllability and observability conditions. Nevertheless, note that \cite{al2017two}, \cite{wu2019multiemitter} don't take a channel into account (only noise), and therefore the accuracy level is much better than our models, yet not realistic. On the downside, both studies do not require prior information for data driven method as this research method does.
}\textcolor{black}{For an empirical study, a recent study \cite{owfi2023meta} is an experimental study \textcolor{black}{that} makes use of RSS and channel state information for positioning, and states the new trainable model requirement for new environment in the experiments, while the proposed model on this study has \textcolor{black}{succeeded} within a single meta-model. In line with the previously discussed meta-model configuration in Section IV-A-3, a beam prediction application with a comparable MAML is conducted as presented in \cite{yang2022meta}. It is noteworthy that the accuracy reported in \cite{yang2022meta} falls below 60\%, whereas our study achieves a higher accuracy of 75.8\%.
}

\begin{figure}[b] 
    \includegraphics[width=0.51\textwidth]{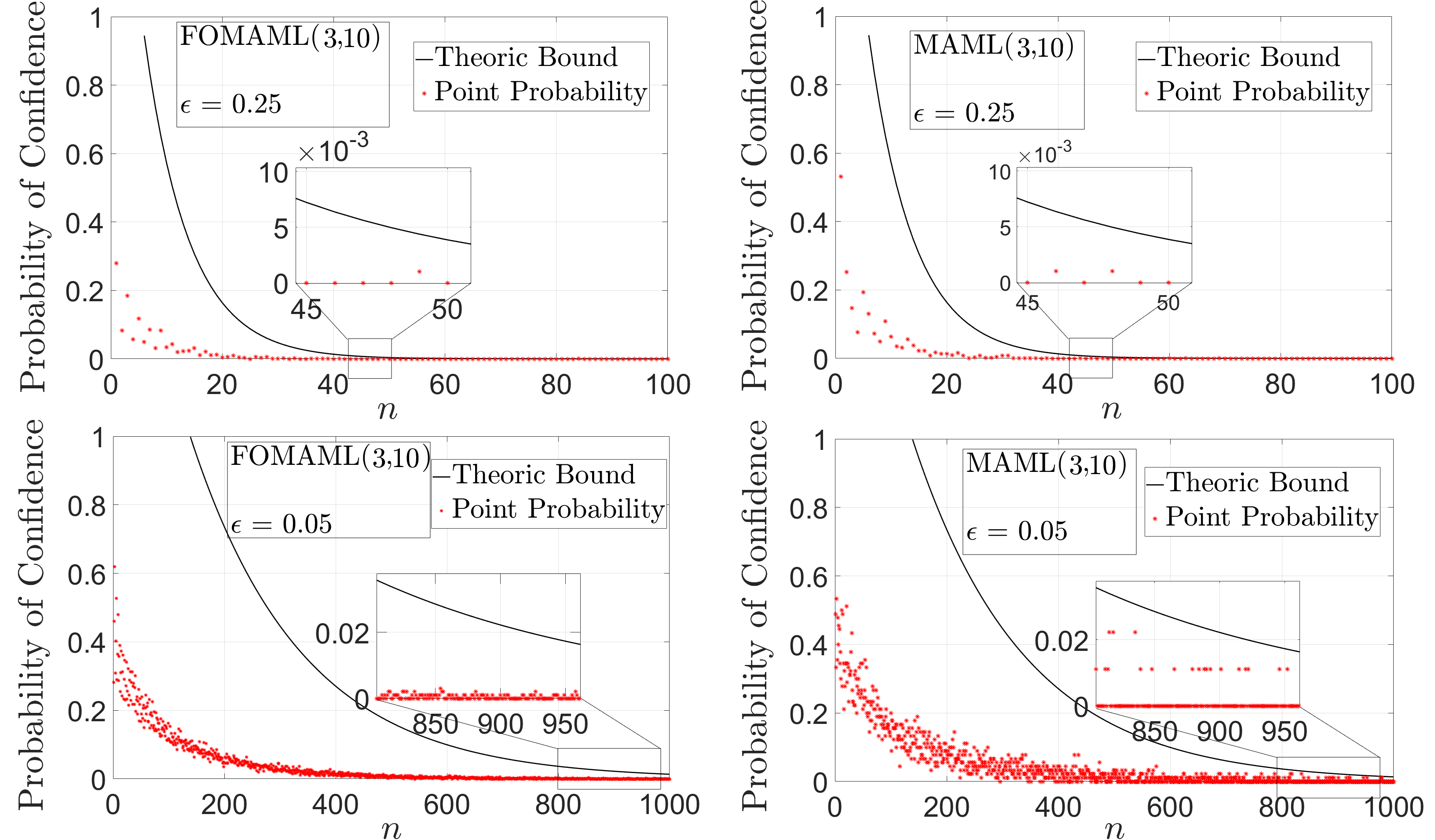}
    \caption{Probability of confidence bounds for both MAML and FOMAML with $\epsilon=0.25$ and $\epsilon=0.05$.}
    \label{hoeffres}
\end{figure}

\textcolor{black}{The performance evaluation revealed that while the FOMAML technique exhibited faster learning capabilities, it also demonstrated slightly lower accuracy than MAML, as expected.} The FOMAML accuracy with $N=10$, $K=3$ is limited with $0.39$ in high SNR whereas the MAML can reach up to $0.57$ accuracy with the same classification scheme. Hence, it can be inferred that the exclusion of the second derivative results in a loss of information in the learning process using 3DPM dataset, in contrast to the general acceptance that FOMAML accuracy is nearly same as MAML. On the other hand, FOMAML is still applicable for its low-cost efficiency since it is able to classify the limited data while CNN cannot. On top of this, FOMAML azimuth accuracy \textcolor{black}{is} directly proportional \textcolor{black}{to} the FOMAML elevation accuracy, as MAML technique does.

On the Figure \ref{hoeffres}, hereusticly, PoC bound indicates the performance difference between MAML and FOMAML where $n>45$ approaches for point probability in $\epsilon=0.25$ and $n>500$ for point probability in $\epsilon=0.05$. Point probabilities show that MAML is more robust in strict classification than FOMAML in low $\epsilon$ thresholds at the same $n$. \textcolor{black}{
Figure \ref{hoeffres}(a) and \ref{hoeffres}(b) depicts the $\epsilon=0.25$ tolerance cases for MAML and FOMAML. It can be seen that, with such a high tolerance level, the amount of sample required is nearly similar, while MAML convergence slightly better at the same $n$, as such \textcolor{black}{highlighted} interval of $45-50$. This convergence gap is even wider in \textcolor{black}{the} lower tolerance threshold of $\epsilon = 0.05$ as can be seen \textcolor{black}{in} Figure \ref{hoeffres}(c) and Figure \ref{hoeffres}(d). Comparing these with the first two subfigures in a and b, the required $n$ is substantially increased and the convergence points are not as close. On this basis, it can be concluded that for 10 times more samples, the error tolerance for MAML(3,10), FOMAML(3,10) applications can be reduced from 0.25 to 0.05. Similarly, for more sensitive applications MAML offers a much safer option, while for low tolerance applications FOMAML can also be viable.}

In terms of the impact of 3DPM, it has been noted that $d_x, d_y, d_z$ related $\delta_{T1}$ and $\delta_{R1}$ results with coherency problem between \textcolor{black}{the} dataset and the model as the misalignment. Nevertheless, this drawback can be suppressed by lowering the $N$, the beam sampled classes, in return \textcolor{black}{for} accuracy performance. \textcolor{black}{
It has been observed that the accuracy levels \textcolor{black}{go} down to 30\% in the Rayleigh fading case evaluation environment. This performance is notably lower compared to LoS scenarios. In NLoS case, the RSS presents low dimension information to separate such a complex scenario, this environment challenges the purpose we try to accomplish.} 

\textcolor{black}{
A significant contribution of the proposed angle estimation method is, that the UAVs don't require mutual information of each other (or require state estimation). Meta-model estimator requires the use of only a data driven model and a real time RSS. Note that the necessary information may be required as the system deviates from the ideal state. During the real time RSS processing, the age of information of RSS is directly related to the meta-model performance. Similarly, a power optimization over the network also negative impact over the proposed estimator for the designed 3DPM dataset. As the number of effects on the system model is increased, our model becomes gradually unstable, just as any machine learning approaches. On the upside, with more prior information, the system can be even more generalized.
}

 \textcolor{black}{
The limitations of the proposed model's accuracy are mainly the due to the selection of limited sized dataset. In the realm of AI's progression towards leveraging small datasets, we aim to showcase the efficacy of meta-learning intentionally designed for limited datasets. Meta-learning proves to be an ideal tool for this purpose, leveraging subdatasets instead of the data itself. Emphasizing that the quality and diversity of the data hold greater significance than sheer volume, we acknowledge that dealing with limited data poses challenges such as overfitting, particularly evident in the case of CNN. \textcolor{black}{Interestingly, employing DenseNet within the framework of MAML rather than a simple CNN presents a promising avenue for enhancing the precision of incident wave angle estimation, albeit at the cost of increased computational complexity.} Regarding FOMAML, its influence manifests diversely across various types of studies. The authors on \cite{rajeswaran2019meta} demonstrate that experiments on MiniImageNet and Omniglot reveal that FOMAML performs almost as well as MAML. However, this outcome is not universally guaranteed, which leads to alternative models such as Reptile \cite{nichol2018first}. In our model, we observe the rapid convergence of FOMAML, consistent with its claims. However, it's worth noting that the significance of the second gradient loss for the given task is substantial, leading to a noticeable reduction in accuracy.
}

\begin{table}[]
  \centering
  \caption{Accuracy performance of the models in 10dB}
  \resizebox{0.75\columnwidth}{!}{%
    \begin{tabular}{c|c|ccc}
    Method                &        & \multicolumn{1}{c}{$N=6$ } & \multicolumn{1}{c}{$N=8$} & {$N=10$} \\ \hline \hline
    \multirow{3}{*}{MAML} & $K=1$ & \multicolumn{1}{c|}{0.443} & \multicolumn{1}{c|}{0.364} & 0.313  \\ 
                          & $K=3$ & \multicolumn{1}{c|}{0.754} & \multicolumn{1}{c|}{0.547} & 0.515  \\  
                          & $K=5$ & \multicolumn{1}{c|}{-}     & \multicolumn{1}{c|}{0.700} & 0.649  \\ \hline
    FOMAML                & $K=3$ & \multicolumn{1}{c|}{-}     & \multicolumn{1}{c|}{-}     & 0.272  \\ \hline
    CNN                   & -      & \multicolumn{3}{c}{0.16  $\pm$0.12} \\ \hline
    DenseNet                   & -      & \multicolumn{3}{c}{0.23 $\pm$0.12}   \\ \hline \hline
    \end{tabular}
    }
    \label{tab4}
\end{table}



\section{Recommended Research Directions}
\label{secopen}
Extracting the incident wave angle by limited RSS information in non-ideal A2A networks is a big step for both communication and sensing applications. It enables the followings that can be studied in further detail,
\begin{itemize}
    \item Demonstrates global observability to assure controllability on non-linear dynamic systems,
    \item Eliminates the requirement for the position information of the mutual, and make relative position estimation possible in radio localization systems,
    \item Keeps the connectivity by lowering antenna impact on the channel without external aid (i.e. intelligent surfaces), resulting with lower packet outages,
    \item Maximizes the LoS capacity for mobile A2A communication,
    \item Eliminates the requirement of the orientation and position information for the sake of sustainable physical channel,
    \item Provides secure communication by detecting unauthorized incident wave from unexpected angles,
    \item Adapts the A2A network for wide range of channel fading distributions in directional communication in THz band by beamforming and beamfocusing,
    \item Decreases the co-channel interference and electromagnetic interference by increasing the radiation efficiency, 
    \item \textcolor{black}{
   Enables the use of different data driven models for wave angle estimation with different application needs.
}
\end{itemize}

\section{Conclusion}        
\label{sec7con}

In conclusion, a thorough 3D geometry based A2A link under \textcolor{black}{a} dynamic antenna with alignment, field pattern and polarization imperfections \textcolor{black}{has} been structured. Under these antenna imperfections, RSS \textcolor{black}{has} been exploited to estimate incident wave angles by using the L2L concept of MAML that can be generalized for any \textcolor{black}{A2A} fading channel with less than $10^5$ limited amount of data. Going one step further, FOMAML with less computation requirement has been deployed to cross the limits of meta-optimization for incident wave angle estimation, utilizing only first order gradient. An upper bound for the MAML accuracy have been established to compare several case demonstrations. It has been observed that radiation intensity on a fading channel has \textcolor{black}{a} vital impact on 3D perspective, even if the radiator is an ideal omnidirectional antenna. MAML can reach up to $0.85$ training accuracy and $0.754$ incident elevation angle accuracy $0.722$ incident azimuth angle accuracy under the mentioned antenna imperfections, whilst CNN barely \textcolor{black}{converges} with $0.16$ incident elevation angle accuracy using same dataset. On the other hand, $\mathrm{DenseNet}$ menages the learning with reaching up to $40\%$ in $\mathrm{DenseNet}-5$, yet still fails in the evaluation for unseen tasks. Furthermore, FOMAML has the fastest convergence rate for limited iteration, yet it fails to reach MAML capabilities with the same shot and class configuration. Lastly, we present our vision \textcolor{black}{of} the potential usage of \textcolor{black}{the} proposed end-to-end model.

\section*{Appendix A}
The radiation power density $P_r=\frac{1}{2}\mathbb{R}[\textbf{E}\times \textbf{H}^*]$ and the radiation intensity $U$ relation is following
\begin{equation}
    P_r  = \frac{1}{d^2} U(\psi,\theta), \textrm{W/m}^2
\end{equation}
The superimposed electrical field ($\textbf{E}(r,\theta,\psi))$) components are following 
\begin{align}
&E_\theta \simeq j\eta \frac{k_w I_0 Le^{-jkd}}{4\pi d} \sin{\theta},\\
&E_r \simeq E_\psi = 0,
\end{align}
where $\eta$ is intrinsic impedance and $k_w = \frac{2 \pi}{\lambda}$ is the wave number for the antenna with $I_0$ current density. Note that due to the far-field assumption, the relation between $\textbf{E}$ and $\textbf{H}$ is linear with $\eta$. Therefore, $U=\frac{d^2}{2\eta}|\textbf{E}|^2$ is a function of $E_\theta$ solely, where the $\mathbb{E}(E_\theta) \simeq \sin{\theta}$. The updated radiation density is the following  
\begin{align}
    &U(\theta) = \frac{d^2}{2\eta}|E_\theta|^2 = \eta \frac{k_w^2 (I_0 L)^2}{32\pi^2} \sin^2{\theta}\\
    &\mathbb{E}(U(\theta))= \sin^2{\theta}
\end{align}
Therefore, the directivity of the antenna ($D=\frac{\max U}{U_0}$) with $\phi^{\psi,\theta}_{T,R}$ is following
\begin{equation}
    D(\psi,\theta) = 4 \pi \frac{\max U(\psi,\theta)}{P_r} = 4 \pi \frac{\max U(\theta)}{P_r}. 
\end{equation}
where $U_0$ is the isotropic source radiation density.

\bibliographystyle{IEEEtran}
\bibliography{reference}
\newpage
\begin{IEEEbiography}[{\includegraphics[width=1.10in,keepaspectratio]{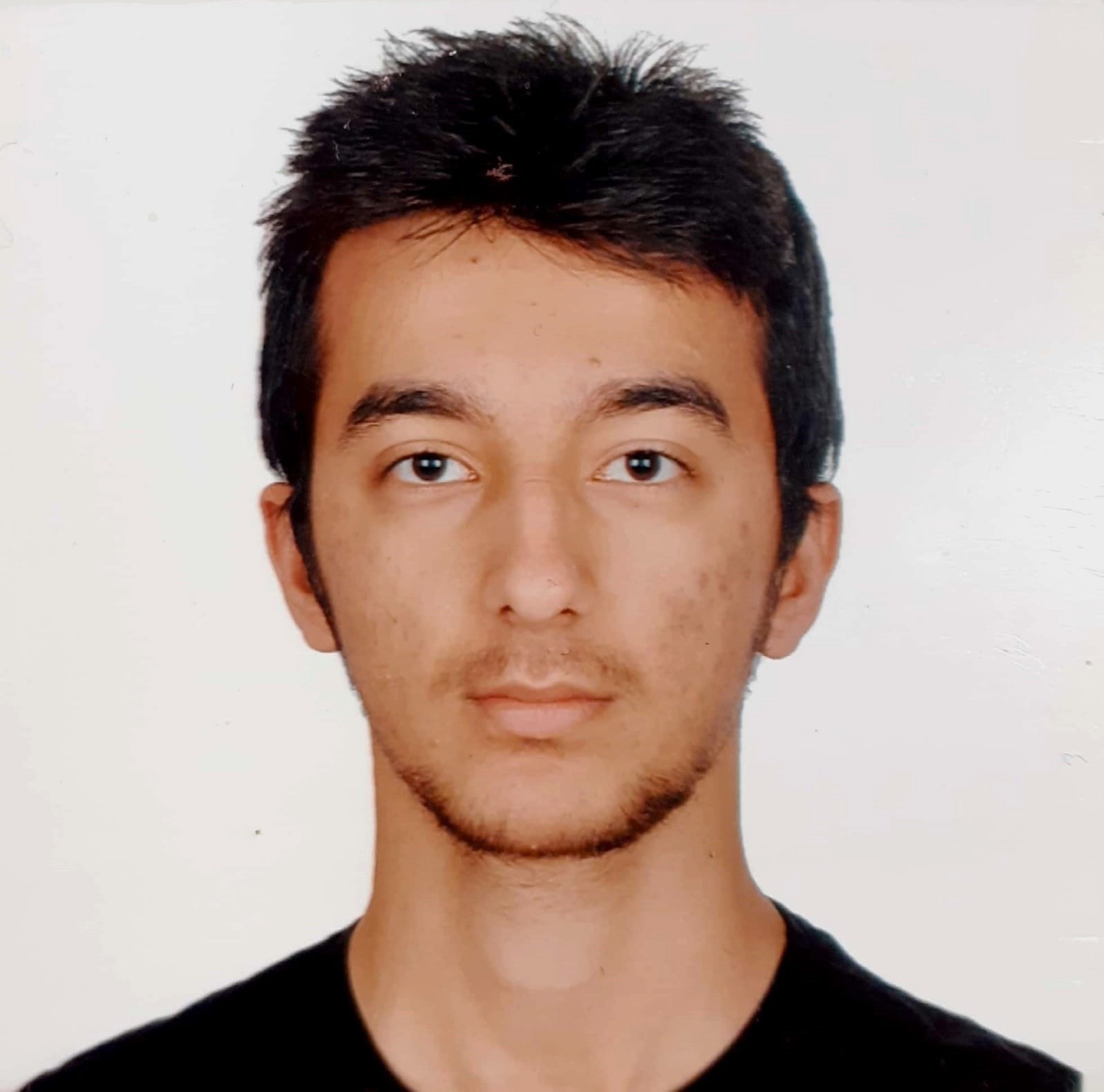}}]%
{\\Eray Güven}
(guven.eray@polymtl.ca) received his B.S. degree in Electronics and Communication Engineering at Istanbul Technical University, Turkey in 2021. He is currently studying for a Ph.D. degree in electrical engineering at Polytechnique Montréal, Montréal, Canada.
\end{IEEEbiography}
\vskip 0pt plus -1fil
\begin{IEEEbiography}[{\includegraphics[width=1.10in,keepaspectratio]{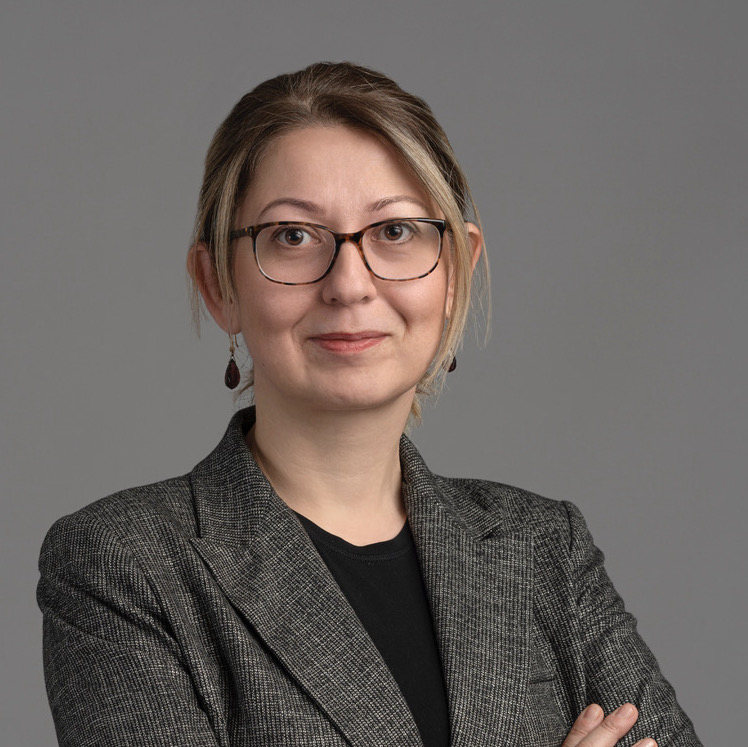}}]%
{\\Güneş Karabulut Kurt}
(gunes.kurt@polymtl.ca), IEEE Senior Member, is a Canada Research Chair (Tier 1) in New Frontiers in Space Communications and Associate Professor at Polytechnique Montréal, Montréal, QC, Canada. From 2010 to 2021, she was a professor at Istanbul Technical University. She is a Marie Curie Fellow and has received the Turkish Academy of Sciences Outstanding Young Scientist (TÜBA-GEBIP) Award in 2019. She received her Ph.D. degree in electrical engineering from the University of Ottawa, ON, Canada. Her research interests include multi-functional space networks, space security, and wireless testbeds.
\end{IEEEbiography}

\end{document}